\documentclass[prb,twocolumn,preprintnumbers,showpacs,amsmath,amssymb]{revtex4}

\usepackage{graphicx}
\usepackage{dcolumn}
\usepackage{bm}

\begin{document}

\input epsf.sty



\title{Interpretation of low-temperature nuclear quadrupole resonance spectra in La$_{1.875}$Ba$_{0.125}$CuO$_4$ in terms of two-dimensional spin superstructure}

\author{Boris V. Fine}

\affiliation{ Department of Physics and Astronomy, University of Tennessee, 101 South College, 1413 Circle Dr., Knoxville, TN 37996, USA}

\date{May 5, 2006}

\begin{abstract}
This paper reanalyzes the low temperature nuclear quadrupole resonance (NQR) experiments in La$_{1.875}$Ba$_{0.125}$CuO$_4$ by 
Hunt {\it et al}. [Phys. Rev. B 64, 134565 (2001)] in an attempt to determine the dimensionality of spin modulations in this and other compounds of the lanthanum family of high temperature cuprate superconductors. It is concluded that the shape of the NQR spectra obtained by Hunt {\it et al}. favors the two-dimensional pattern of spin modulations known as ``grid'' or ``checkerboard''.  The paper also contains the discussion of charge patterns, which can accompany the above spin modulation.
\end{abstract}
\pacs{}


\maketitle

\narrowtext
\pagebreak

\section{Introduction}
\label{intro}

\begin{figure} \setlength{\unitlength}{0.1cm}
\begin{picture}(75, 75) 
{ 
\put(1, 0){ \epsfxsize= 2.7in \epsfbox{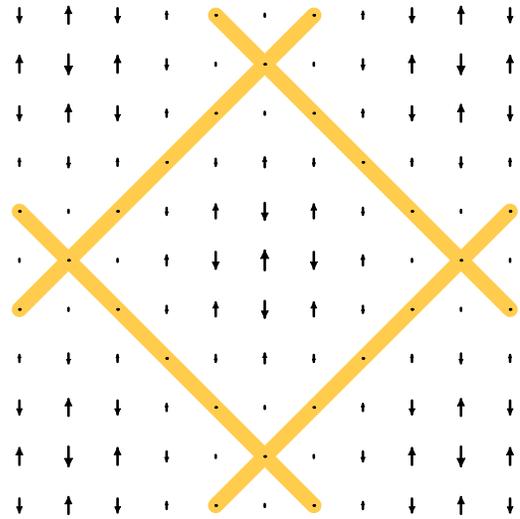  } }
}
\end{picture} 
\caption{[Color online] Commensurate diagonal grid modulation [cite-centered]. The antiferromagnetic order is modulated by function (\ref{f}). The length of the arrows is proportional to the value of staggered magnetization.
Lines drawn through the zero magnetization sites indicate antiphase boundaries of magnetic domains. They form a "grid" pattern.
} 
\label{fig-grid} 
\end{figure}

The notion of stripes has become a standard concept in the field of high-temperature superconductivity about ten years ago, after Tranquada {\it et al}.\cite{Tranquada-etal-95} have observed the four-fold splitting of the antiferromagnetic (AF) $(\pi, \pi)$ peak in {\it elastic} neutron scattering experiments in the lanthanum family of cuprates. The above authors have interpreted the resulting four peaks, as coming from  two different kinds of stripe domains, each exhibiting
a unidirectional  spin modulation (referred to below as 1D) running along one of the two principal lattice directions. The 1D spin modulation was to be accompanied by the 1D charge modulation peaked around the antiphase magnetic boundaries, hence the name ``stripes". There exists, however, an alternative interpretation sometimes referred to as grid or checkerboard, which stipulates that the four-fold splitting of the $(\pi, \pi)$ neutron peak is caused by a two-dimensional (2D) modulation of the AF order shown in Fig.~\ref{fig-grid}. Such a modulation is accompanied by a 2D charge density wave localized around the antiphase boundaries, which have the appearance of grid --- the term to be used in the rest of this paper. 


Grid superstructure has been mentioned in the literature since the early numerical works on the subject\cite{ZG,Kato-etal-90,Tranquada-etal-99,SG}, however, it was only quite recently\cite{Fine-hitc-prb04,Fine-Miami-04}, that grid was advocated as a preferred interpretation of the $(\pi, \pi)$ peak splitting. 
It has also been suggested in Ref.\cite{Fine-hitc-prb04} that similar 2D inhomogeneous structures are universally present in other families of cuprates in the normal and superconducting states. The insight of that work was not completely trivial, since its appearance predated the scanning tunneling spectroscopy (STS) studies of Vershinin {\it et al}.\cite{Vershinin-etal-04}, Hanaguri {\it et al}.\cite{Hanaguri-etal-04},
and the neutron study of Hinkov {\it et al}.\cite{Hinkov-etal-04} --- all three are the works, that brought the dimensionality of generic spin modulations in cuprates back into the spotlight (see, e.g.,\cite{Yao-etal-06}).

The connection between the inhomogeneous structures in lanthanum cuprates and the generic properties of other high-temperature cuprate superconductors is a delicate issue. Spin and charge modulations observed in this cuprate family may turn out to be a composition-specific feature, or, in the opposite extreme, similar modulations may be universally present in all families of cuprates but observable only in the lanthanum family due to anomalous slowing down. In this context, the difference in the dimensionality of the modulation is not inconsequential.
The apparent message contained in the symmetry of the superconducting order parameter\cite{AGL}, the symmetry of the pseudogap\cite{DHS}, the STS observations on the surface of Bi$_2$Sr$_2$CaCu$_2$O$_{8+\delta}$ (Bi-2212)\cite{Hoffman-etal-02,Hoffman-etal-02A,Howald-etal-03,Vershinin-etal-04} and Ca$_{2-x}$Na$_x$CuO$_2$Cl$_2$\cite{Hanaguri-etal-04} is two-dimensional.  Therefore, if the 1D nature of the spin modulations in lanthanum cuprates is established conclusively, then, in the opinion of this author, it will likely imply the composition-specific nature of the phenomenon. (The authors of Refs.\cite{Kivelson-etal-03,Robertson-etal-06} seem to have the opposite expectation.) 
In contrast, if the spin modulations in this cuprate family are two-dimensional, then this would be yet another indication that the tendency to form 2D modulations is a general property of cuprates and, therefore, may be essential for the mechanism of superconductivity.
The differences between different geometries of 2D  modulations should be much less important. In lanthanum cuprates, magnetic modulations may develop into a diagonal grid pattern, while, in other cuprates, the system may phase separate into more-AF-correlated and less-AF-correlated nano-clusters\cite{Stock-etal-06} without forming quasi-static magnetic order and, therefore, not accompanied by antiphase boundaries.

Since the argument for the non-universality of the modulations in lanthanum cuprates has merit, it is important to establish the dimensionality of these modulations without invoking the knowledge about other families of cuprates or nickelates. The discussion in the literature specific to lanthanum cuprates has been limited so far mostly to neutron\cite{Tranquada-etal-95,Tranquada-etal-96,Yamada-etal-98,Tranquada-etal-99,Fujita-etal-02,Fujita-etal-04} and hard X-ray\cite{Zimmermann-etal-98} experiments.
The primary focus of this paper is on extracting the dimensionality of spin modulations in La$_{1.875}$Ba$_{0.125}$CuO$_4$ from the nuclear quadrupole resonance (NQR) results of Hunt {\it et al}.\cite{Hunt-etal-01} in combination with the muon spin rotation ($\mu$SR) results of Nachumi {\it et al}.\cite{Nachumi-etal-98}. The paper also elaborates on some of the dimensionality-related arguments made earlier in Ref.\cite{Fine-hitc-prb04}. Most of the paper will  contrast 1D stripes with 2D grid, but 1D spin spirals\cite{Shraiman-etal-88,Hasselmann-etal-04,Sushkov-etal-05} will also be discussed in Section~\ref{discussion}. 

The plan of the rest of the paper is as follows.
In Section~\ref{review}, the present non-NQR knowledge about the dimensionality of spin and charge modulations in lanthanum cuprates is discussed. Section~\ref{previous} discusses relevant NQR findings of Hunt {\it et al}. The NQR-related findings of the present work are presented in Section~\ref{computational} and discussed in Section~\ref{discussion}. 

Sections \ref{review} and \ref{previous} amount [mainly] to an extended introductiry review. The readers familiar with the broader issues involved can skip the above sections and proceed directly to Section~\ref{computational}.

\section{Charge modulations in the lanthanum family of cuprates}
\label{review}

Magnetic peaks corresponding to higher order modulation harmonics have not been observed so far and thus cannot be used to discriminate between different modulation patterns.  A direct resonant X-ray observation of charge carrier density modulations reported  recently by Abbamonte {\it et al}.\cite{Abbamonte-etal-06} has not addressed the dimensionality issue either.
The direct attempts to discriminate between 1D stripe and 2D grid patterns have been made so far only on the basis of the orientation of charge satellites\cite{Tranquada-etal-99,Kivelson-etal-03,Robertson-etal-06}, which represent the response of the lattice to the periodic charge and/or spin modulations. 

\begin{figure} \setlength{\unitlength}{0.1cm}
\begin{picture}(75, 210) 
{
\put(-5, -5){ \epsfxsize= 2.6in \epsfbox{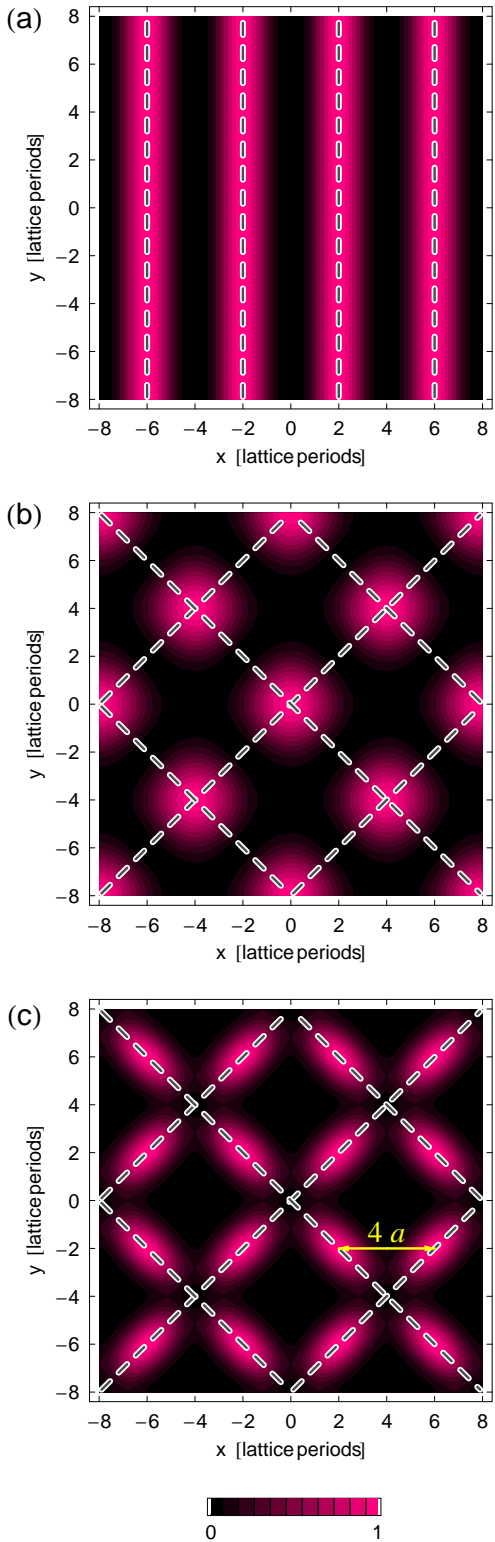  } }
}
\end{picture} 
\caption{[Color online] Three color-coded patterns of charge modulations corresponding to the same splitting of magnetic $(\pi,\pi)$ peak: (a) stripes; (b) grid in the Landau approximation; (c) grid with the charge density avoiding the grid nodes [Eq.(\ref{rho})]. Dashed lines represent antiphase boundaries of antiferromagnetic domains.
} 
\label{fig-patterns} 
\end{figure}

In Refs.\cite{Kivelson-etal-03,Robertson-etal-06}, the analysis of experiments is based on Landau expansion\cite{ZKE} , which requires small spin and charge modulations. The resulting charge patterns for the 1D and the 2D cases are shown in Fig.~\ref{fig-patterns}(a,b).  In the framework of Landau expansion, leading charge peaks corresponding to 1D stripe and 2D grid pictures do not coincide, and, therefore,  the observation of charge satellites expected for the 1D case together with the non-observation of the satellites expected for the 2D case (notwithstanding the weakness of the charge signal) do present a clear argument. 


There are, however, good reasons to expect that the amplitude of the spin modulations is not small. For example, as indicated by $\mu$SR\cite{Nachumi-etal-98}, the amplitude of the modulated staggered magnetization in La$_{1.875}$Ba$_{0.125}$CuO$_4$ is large --- about half of the staggered moment of the undoped parent compound. Strong modulation of staggered magnetization would induce a deeper local potential for charge carriers thus forcing them to be stronger localized around the antiphase boundaries. Larger amplitude of charge modulation implies stronger interaction between the charge carriers, which would undermine the theoretical basis for the Landau expansion and introduce new charge peaks. One particular problem with the Landau expansion is that it gives the maximum of charge density around the nodes of the antiphase grid [see Fig.~\ref{fig-patterns}(b)].
If the charge modulations are large, then the Coulomb repulsion between the charge carriers would rather lead to a smaller charge density around the nodes.

In the case of large charge modulations, the leading charge peaks corresponding to the 1D and 2D interpretations of the magnetic $(\pi, \pi)$ peak splitting can coincide.  If charge density is reasonably localized around the antiphase grid but avoids grid nodes, then a useful insight can be gained from analysing the density profile shown in Fig.~\ref{fig-patterns}(c). It is given by function
\begin{eqnarray}
\nonumber
\rho(x,y) &=& \left[ \hbox{cos} \frac{\pi(x+y)}{8 a} \right]^8 \  
\left[ \hbox{sin} \frac{\pi(-x+y)}{8 a} \right]^2
\\ 
&& + \ \left[ \hbox{cos} \frac{\pi(-x+y)}{8 a} \right]^8 \ 
\left[ \hbox{sin} \frac{\pi(x+y)}{8 a}\right]^2 ,
\label{rho}
\end{eqnarray}
where $x$ and $y$ are the coordinates along two principal lattice directions and $a$ is the lattice period.
This profile was mentioned in Ref.\cite{Fine-hitc-prb04} but not elaborated upon.   It corresponds to the same splitting of magnetic $(\frac{\pi}{a}, \frac{\pi}{a})$ peak as the one arising from the 1D stripes spaced by $4a$. The cos$^8$ factors in Eq.(\ref{rho}) set the grid pattern, while the sin$^2$ factors suppress the intensity near the grid nodes. The positions of the resulting charge peaks and their relative intensities are shown in Fig.~\ref{fig-peaks}. (See the Appendix for the calculation.)  Leading charge peaks corresponding to 1D stripes spaced by $4a$ and to modulation (\ref{rho})   coincide with each other and are located at 
$(\pm \frac{\pi}{2 a}, 0) $ and $(0, \pm \frac{\pi}{2 a})$, where charge peaks were detected in Refs.\cite{Tranquada-etal-95,Zimmermann-etal-98,Tranquada-etal-99,Fujita-etal-02,Fujita-etal-04}.  Modulation (\ref{rho}) also generates peaks at $(\pm \frac{\pi}{4 a}, \pm \frac{\pi}{4 a}) $ predicted by Landau expansion, but these peaks have only third largest intensity (about 7 times smaller than the intensities of the leading peaks). The peaks of the second largest intensity (4 times smaller than that of the leading peaks) are located at 
$(\pm \frac{\pi}{2 a}, \pm \frac{\pi}{2 a}) $. 

It may appear that the leading peak position for the density profile (\ref{rho}) is the result of a fine tuning of the powers of sine and cosine. This is, however, not the case. The only important aspect of this profile is that the factors sin$^2$ suppress the charge density near the grid nodes, as required by the Coulomb repulsion between charges. As illustrated in the Appendix, if the sin$^2$ factors are removed, then the leading peaks are located at the position predicted by the Landau expansion, but, if they are left, then the position of the leading peaks at 
$(\pm \frac{\pi}{2 a}, 0) $ and $(0, \pm \frac{\pi}{2 a})$ 
does not depend on the power of the cosine factors.  The unrealistic aspect of profile (\ref{rho}) is that the sin$^2$ factors suppress the charge density near the grid nodes {\it completely}. However, even an incomplete Coulomb suppression of the charge density near the grid nodes can easily shift the intensity maximum from the Landau expansion result to $(\pm \frac{\pi}{2 a}, 0) $ and $(0, \pm \frac{\pi}{2 a})$.


\begin{figure} \setlength{\unitlength}{0.1cm}
\begin{picture}(75, 75) 
{ 
\put(1, 0){ \epsfxsize= 2.7in \epsfbox{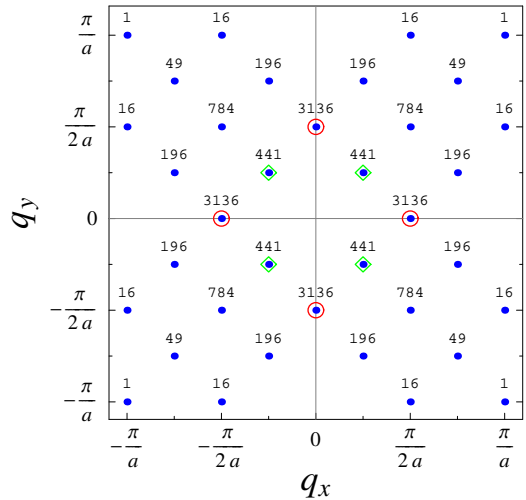  } }
}
\end{picture} 
\caption{[Color online] Positions of charge peaks (filled circles) corresponding to the scattering from the charge modulation given by Eq.(\ref{rho}) and shown in Fig.~\ref{fig-patterns}(c). Numbers above the circles represent the relative intensity of the peaks obtained in the Appendix. Large empty circles and empty diamonds are the positions of leading charge peaks corresponding, respectively, to stripes [Fig.~\ref{fig-patterns}(a)] and to grid in the Landau approximation [Fig.~\ref{fig-patterns}(b)] 
} 
\label{fig-peaks} 
\end{figure}


From the above perspective, it is the non-observation of the weaker $(\pm \frac{\pi}{4 a}, \pm \frac{\pi}{4 a}) $ and $(\pm \frac{\pi}{2 a}, \pm \frac{\pi}{2 a}) $ peaks rather than the observation of the $(0, \pm \frac{\pi}{2 a})$ peaks that would be the real argument in favor of  the 1D interpretation. However, to the best of this author's knowledge, the open literature contains only one experimental scan across the $(\frac{\pi}{4 a},  \frac{\pi}{4 a}) $ and $( \frac{\pi}{2 a},  \frac{\pi}{2 a}) $, which has been presented in Ref.~\cite{Tranquada-etal-99} for La$_{1.48}$Nd$_{0.4}$Sr$_{0.12}$CuO$_4$. In that case, the experimental resolution was just not sufficient to claim observation or non-observation of the peaks, which would be 4 or 7 times weaker than the leading peak observed at $(0, \frac{\pi}{2 a})$. In fact, quite remarkably the scan around 
$( \frac{\pi}{2 a},  \frac{\pi}{2 a})$ shows a promising sign of a peak. Repeating that scan with a better resolution would certainly have merit.

Another argument frequently cited in support of the 1D stripe picture is that the observability of 
static spin and charge modulations is correlated with the onset of the ``non-generic'' low temperature tetragonal (LTT) crystal phase. Tranquada {\it et al}.\cite{Tranquada-etal-95} have proposed, that the 1D stripes follow the lattice anisotropy of the LTT phase. This anisotropy alternates between adjacent CuO$_2$ planes, which would imply, that stripes in two adjacent planes run perpendicular to each other. 
Empirically, however, the onset of the LTT phase is not a necessary condition for the observability of static modulations, because quasielastic signatures of spin modulations have also been reported for the ``generic'' low temperature orthorhombic (LTO) phase of La$_{2-x}$Sr$_x$CuO$_4$ with and without Zn codoping\cite{Hirota-etal-98,Kimura-etal-99,Hirota-01}. 
But one can still hope that the presence of the LTT phase is at least a sufficient condition for the 
observability of static modulations.

Although the above proposition of Tranquada {\it et al}. has visual appeal, the onset of the LTT phase as such does not readily discriminate between the stripe and the grid interpretations. The general expectation is that the magnetic $(\pi, \pi)$ peak splittings observed by elastic neutron scattering in the LTT phase and by inelastic neutron scattering in the LTO phase have the same origin. If true, this implies that the system has chosen the dimensionality of the modulation pattern (1D or 2D) without the LTT phase being involved, and then the LTT phase only plays a role in slowing the modulations down, to make them observable in elastic experiments\cite{LTTenergy}. Therefore, to claim that the symmetry of the LTT phase discriminates between the grid and the stripe pictures is equivalent to claiming that the LTT phase would slow down fluctuating stripes, but would {\it not} slow down fluctuating grid. This author is not aware of any experimental or theoretical result, which would support the latter part of this claim.

The picture of stripes running in perpendicular directions in the adjacent planes has further difficulty that such a configuration is not most favorable energetically from the viewpoint of interplane Coulomb interaction. The lowest Coulomb energy corresponds to stripes  running in the same direction in two adjacent planes with pattern in one plane shifted with respect to the other by half a period of charge modulation.  Since the stripes are unscreened lines of charge, the interplane Coulomb energy should be large. Can the coupling of stripes with the lattice anisotropy outweigh the Coulomb coupling between stripes from adjacent planes? It does not seem very likely.

Recently, Robertson et al.\cite{Robertson-etal-06} have also argued that, due to the anisotropy of the LTT phase, the incommensurate grid modulation should have slightly different periodicity in the adjacent CuO$_2$ planes, which should lead to 8-fold splitting of the $(\pi, \pi)$ peak instead of the 4-fold splitting observed experimentally. Since the crystal lattice periods cannot be detectably different in the adjacent planes (because of the prohibitive elastic energy cost), the argument of Roberson et al. is not very direct, as it relies on the additional theoretical assumption, that the mechanism similar to the Fermi-surface nesting is behind the formation of the grid superstructure. This argument has further difficulty, that the mismatch of the charge superstructure periods would lead to the regions, where the maxima of charge modulation density in two adjacent planes lie on the top of each other, which, in turn, leads to a large Coulomb energy cost.

The critical discussion so far does not change the fact that the observed positions of the charge peaks, and the involvement of the LTO-LTT transition
cannot be turned into an argument favoring the grid over the stripe pictures, and to this extent, these two observations do support the 1D stripe interpretation.
In the present case, however, the the presumption of innocence can be rephrased as follows: A two-dimensional experimental peak pattern should be considered as an evidence for two-dimensional superstructure unless conclusively proven otherwise. In the opinion of this author, the above cited facts fall far short of delivering a conclusive proof.

One can, however, start from the ``presumption of guilt" by considering the 2D grid interpretation of the 2D neutron data as ``exotic", and requiring compelling reasons to choose it over the 1D stripe picture. It is the purpose of the present paper to analyse the NQR experiments from this perspective, and to point out that, within the limits of experimental uncertainties, they do constitute a clear indication in favor of the 2D grid interpretation.

\section{Discussion of NQR results by Hunt {\it et al}.}
\label{previous}

\begin{figure} \setlength{\unitlength}{0.1cm}
\begin{picture}(75, 140) 
{
\put(-5, 0){ \epsfxsize= 3.3in \epsfbox{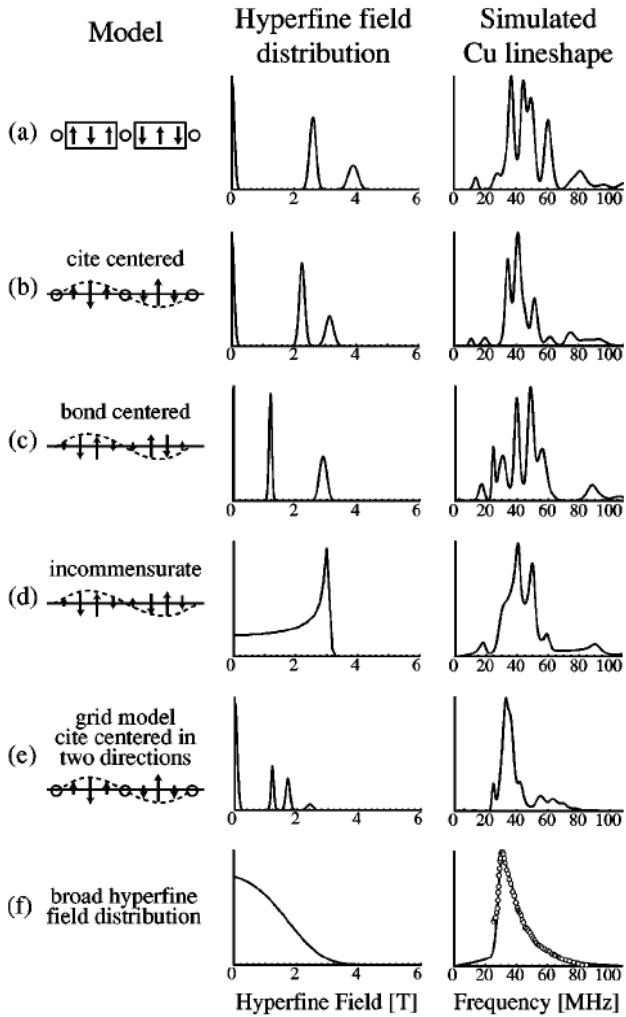  } }
}
\end{picture} 
\caption{[Reproduced from Ref.\cite{Hunt-etal-01}] Distributions of hyperfine fields and corresponding NQR spectra obtained in Ref.\cite{Hunt-etal-01}. Solid lines represent theoretical calculations. Circles in
figure (f) represent experimental NQR spectrum at 350 mK. Vertical axes have arbitrary units everywhere. 
} 
\label{fig-hunt} 
\end{figure}

A slowly fluctuating spin modulation, 1D or 2D, is expected to freeze at sufficiently low temperatures thus leading to a broad NQR spectrum having shape, which may discriminate between different dimensionalities of the modulation. 
In Ref.[1], Hunt {\it et al}. have presented, among other findings, the results of their NQR study of La$_{1.875}$Ba$_{0.125}$CuO$_4$  aimed at testing the stripe interpretation of neutron scattering experiments. 
At temperatures below 8K, they have, indeed observed a broad NQR spectrum of $^{63}$Cu [shown in Fig.~\ref{fig-hunt}(f)], which signified the expected freezing of local hyperfine fields affecting copper nuclear spins. This study was done in the context of a broader effort to elucidate the dynamics of electronic spins in lanthanum cuprates by the means of nuclear magnetic and quadrupole resonances\cite{Hunt-etal-99,Curro-etal-00,Suh-etal-00,Teitelbaum-etal-00,Teitelbaum-etal-00A,Haase-etal-00,Haase-etal-02}


\begin{figure} \setlength{\unitlength}{0.1cm}
\begin{picture}(75, 60) 
{ 
\put(-8, 0){ \epsfxsize= 3.4in \epsfbox{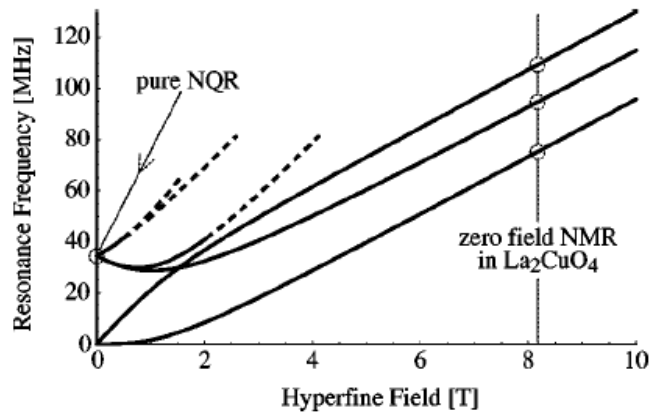  } }
}
\end{picture} 
\caption{[Reproduced from Ref.\cite{Hunt-etal-01}] $^{63}$Cu NQR transition frequencies as a function of hyperfine field. The transition frequencies are obtained from Hamiltonian (\ref{H}). Dashed lines indicate the regions of low probability transitions.
} 
\label{fig-freq} 
\end{figure}

The above NQR spectrum has exhibited one clear feature, the 31~MHz peak, but an important theoretical finding of Hunt {\it et al}. was that this particular peak should be present for any continuous distribution of
hyperfine fields passing the interval 0.5-1.5~T, where the minima of two NQR frequency branches are located - both corresponding to frequency 31~MHz (see Fig.~\ref{fig-freq}). Therefore, the presence of the 31~MHz peak could not discriminate between different kinds of modulations.  Apart from this peak, the spectrum  has exhibited no other structure.


Hunt {\it et al}. have attempted to fit
the NQR spectrum to the predictions of many possible 1D {\it static} stripe pictures, 
but have found that all such pictures lead to very structured lineshapes having several peaks not observed experimentally [see Figs.~\ref{fig-hunt}(a-d)]. 

The origin of the structure in theoretical NQR lineshapes is two-fold. (i) For commensurate
superstructures, there is only a finite number of distinct nuclear sites. The traces of
this discreetness then appear in the NQR lineshape. (ii) For uniform incommensurate 1D modulations,
which contain the infinite number of distinct sites, the peaks in the NQR lineshapes result from a Van Hove singularity in the distribution of 1D-modulated hyperfine fields. This singularity is located at the maximum in the absolute value of the hyperfine fields [see the middle panel of Fig~\ref{fig-hunt}(d)].

A priori, a 2D grid modulation represents a promising alternative to interpret the rather featureless NQR spectrum of Hunt {\it et al}., because, when the grid is commensurate, it contains a larger number of inequivalent nuclear sites, and thus tends to produce more smeared structure in NQR lineshape, and, when the grid is incommensurate, it would {\it not} lead to a Van Hove singularity at the maximum value of 2D modulated hyperfine fields. (It would lead to a singularity around the zero value of hyperfine field, but that singularity is less important. See the next section.)

Hunt {\it et al}. have considered the possibility of grid modulation, but only the commensurate version with the orientation rotated by 45 degrees in comparison to the one expected from the positions of neutron peaks.
In this way, they have, nevertheless, obtained a much better agreement with experiment in comparison with any of the 1D interpretations [see Fig.~\ref{fig-hunt}(e)], which is not very surprising in view of the preceeding discussion. Yet, the grid pattern used by Hunt {\it et al}. generated too much structure, which led them to abandon the grid interpretation all together and to test another one in terms of the random distribution of hypefine fields. The latter interpretation  gave nearly perfect agreement
with the experimental NQR spectrum [Fig.~\ref{fig-hunt}(f)]. The problem with that interpretation is that the random configuration of staggered spin polarizations is inconsistent with the modulations of AF spin structure observed by neutron scattering\cite{Tranquada-etal-95}. Random picture should also lead to a monotonic decay of the $\mu$SR asymmetry signal rather than the oscillations observed in experiment\cite{Nachumi-etal-98}.

The frequency of the $\mu$SR asymmetry oscillations implies that the modulation amplitude of the staggered polarization is about $0.3 \mu_B$\cite{Nachumi-etal-98}, where $\mu_B$ is Bohr magneton. This number imposes a difficult constraint on the interpretations of NQR spectrum. Hunt {\it et al}., have concluded that any kind of static broadening of the hyperfine field distribution corresponding to the modulation amplitude $0.3 \mu_B$ would fail to reproduce their NQR spectrum, because it would introduce too much spectral weight in the higher frequency interval, where much smaller weight is observed experimentally. They have found, however, that
with the modulation amplitude $0.1-0.2 \mu_B$ and a very significant degree of broadening (50-100 per cent of the modulation amplitude), the observed NQR spectra can be reproduced.

Hunt {\it et al}. have suggested that the above discrepancy may be due to spin fluctuations on the NQR timescale of 20 $\mu$s, which are not
seen on the faster $\mu$SR timescale of 0.5$\mu$s. This suggestion, however, runs into the difficulty, that, in order to change the shape of the effective distribution of hyperfine fields, spins should fluctuate on the time scale of inverse NQR frequency ($\sim 0.03 \ \mu$s) or faster. (This, presumably, happens at temperatures higher than 30~K.) The fluctuations on the scale of 20 $\mu$s can lead to the loss of resonant spins, which, in turn, can reduce the integrated signal intensity but should not change the shape of the spectral line. This is consistent with the fact, that the NQR spectra of Hunt {\it et al}. have identical shape but different intensity at 1.7~K and 350~mK. Even if, one insists that the spin fluctuations may affect the observed NQR lineshape despite the above argument, one would have to make a further assumption that these fluctuations remain temperature-independent as the temperature is reduced by a factor of 5 from 1.7~K to 350~K, which is quite unlikely.

The present work has been motivated by the observation that Hunt {\it et al}. did not consider incommensurate grid modulation, which has promise of describing their experiment without difficult assumptions.

\section{NQR spectra for diagonal grid modulation}
\label{computational}

As mentioned in the previous section, Hunt {\it et al}. have considered grid modulation,
which had different orientation,  in comparison to the one that follows from the neutron experiments. Their grid was oriented along the principal lattice directions,
whereas the geometry of the neutron $(\pi, \pi)$ peak splitting implies the diagonal grid orientation\cite{Fine-hitc-prb04} given by the modulation function [leading harmonic]
\begin{equation}
f(x,y) = \hbox{sin}\left[ {\pi \delta \over a} (x+y) \right]  \hbox{sin}\left[ {\pi \delta \over a} (x-y) \right] 
\label{f}
\end{equation}
and shown in Fig.\ref{fig-grid}. Here  $\delta$ is the dimensionless measure of the splitting of the $(\pi, \pi)$ peak. The four peaks are located at $[{\pi \over a}(1 \pm 2\delta), {\pi \over a}]$  and $[{\pi \over a}, {\pi \over a}(1 \pm 2\delta)]$. The experiments indicate\cite{Fujita-etal-04} that $\delta \approx \frac{1}{8}$, which corresponds to the distance $4 a \sqrt{2}$ between two nearest grid nodes.

In this section, three cases are computed theoretically and compared with experiment: (a) commensurate grid corresponding to $\delta = \frac{1}{8}$; (b) uniform incommensurate grid corresponding to $\delta = \frac{1}{8} - \epsilon$, where $\epsilon$ is an irrational number much smaller than $\frac{1}{8}$; and (c) disordered incommensurate grid with  $\delta \approx \frac{1}{8}$. Below, unless explicitly indicated otherwise, the same computational routine and the same values of parameters as those of Hunt {\it et al}.\cite{Hunt-etal-01} have been used.

When AF spin structure is modulated by some function $f(x,y)$, the hyperfine field at position $(x,y)$ is given by
\begin{eqnarray}
H(x,y) & = & \mu_0  \left\{
A_x f(x,y) - B \left[ f(x+a,y) \right. \right.
\label{Hxy}
\\
&&
\left. \left.
+ f(x-a,y) + f(x,y+a) + f(x,y-a) \right]
\right\} ,
\nonumber
\end{eqnarray}
where\cite{Imai-etal-93} $A_x = 38 \ \hbox{kOe}/\mu_B$ and $B = 42 \ \hbox{kOe}/\mu_B$ are the hyperfine coupling coefficients and $\mu_0 = 0.3 \mu_B$ is the maximum value of staggered polarization.
Substituting $f(x,y)$ given by Eq.(\ref{f}), one can obtain explicitly that 
\begin{equation}
H(x,y) = H_m \ 
\hbox{sin}\left[ {\pi \delta \over a} (x+y) \right]  \hbox{sin}\left[ {\pi \delta \over a} (x-y) \right],
\label{Hxy1}
\end{equation}
where
\begin{equation}
H_m = \mu_0 \left[ A - 4 B \hbox{cos}^2 (\pi \delta) \right].
\label{Hm}
\end{equation}

It is assumed that the hyperfine magnetic fields are parallel to the CuO$_2$ planes\cite{Vaknin-etal-87}, and that the quadrupolar part of the Hamiltonian is axially symmetric with respect to the  crystal axis perpendicular to the CuO$_2$ planes\cite{Hunt-etal-01} ($c$-axis). With these assumptions, the full Hamiltonian for one $^{63}$Cu nuclear spin has form
\begin{equation}
{\cal H} = \frac{\nu_Q h}{2} \left\{I_z^2 - \frac{1}{3} I (I+1)\right\}- \gamma_n h {\mathbf H \cdot \mathbf I}
\label{H}
\end{equation}
where $h$ is the Plank constant, $\mathbf I$ is the spin 3/2 operator, $\mathbf H$ is the hyperfine field, $\gamma_n$ is the giromagnetic ratio of $^{63}$Cu nuclei (equal to 11.285 MHz/Tesla ), and  $\nu_Q$ is the quadrupolar parameter having average values\cite{Hunt-etal-01} 34.5~MHz and 41.8~MHz for for the A- and B-lines of the $^{63}$Cu NQR spectra. The relative spectral weight of the B-line with respect to the A-line  was extracted from the high-temperature NQR spectra of Hunt {\it et al}.\cite{Hunt-etal-01} to be 0.07. In the calculation, the Gaussian distribution of $\nu_Q$ around the above mentioned average values was assumed with root-mean-squared deviation $2.8$~MHz\cite{Hunt-etal-01}. 

For a given value of quadrupolar parameter $\nu_Q$ and the hyperfine field $H$, Hamiltonian (\ref{H}) generates four levels and six transition frequencies between those levels (shown in Fig.~\ref{fig-freq}).

The contribution of each transition to the NQR spectrum was weighed by the square of the transition frequency, and then by the isotropic average  of the transition probability
\begin{equation}
P_{n,n^{\prime}} = | \langle n^{\prime} | \mathbf I  \cdot \mathbf H_1 |n \rangle|^2,
\label{Pnn}
\end{equation}
where $\mathbf H_1$ is the radio frequency field ($\sim$100~G)(See though the discussion of this factor in the next section.) 

When the grid structure is commensurate, it generates only a finite discrete set of different hyperfine fields $\{H_i\}$. This discrete distribution was broadened to a set of Gaussians exp$-\frac{(H-H_i)^2}{2 \sigma_i^2}$ with $\sigma_i = 0.04 H_i$\cite{Hunt-etal-01}.

\begin{figure} \setlength{\unitlength}{0.1cm}
\begin{picture}(75, 105) 
{ 
\put(-44, 0){ \epsfxsize= 6.5in \epsfbox{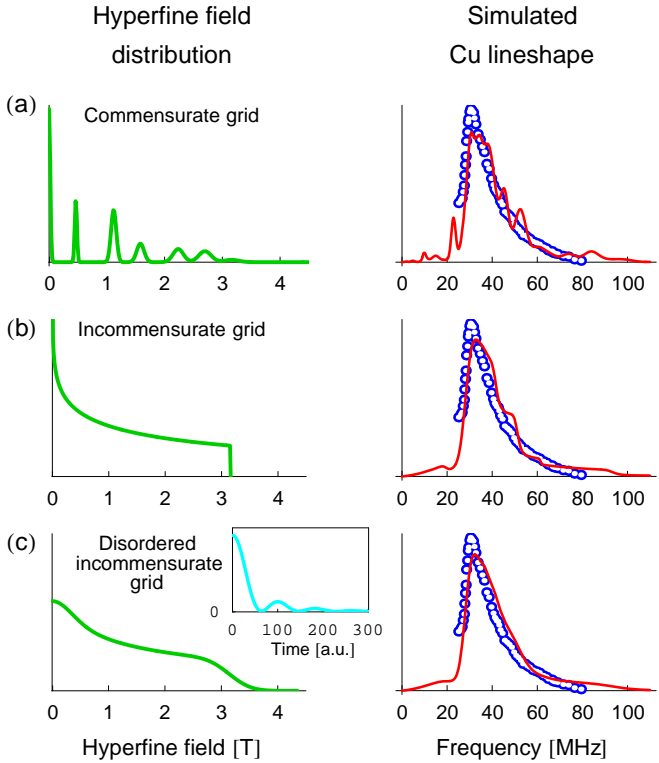  } }
}
\end{picture} 
\caption{[Color online] Distributions of hyperfine fields and NQR spectra corresponding to
(a) commensurate diagonal grid superstructure shown Fig.~\ref{fig-grid};
(b) incommensurate version  of the same grid superstructure, and (c) disorder-broadened modification of case (b). Solid lines represent theoretical calculations. Circles represent 350 mK experimental NQR spectra from Ref.\cite{Hunt-etal-01}. The relative scale of experimental and theoretical NQR spectra is adjusted to give the best agreement between them. The inset in figure (c) shows the Fourier transform of the hyperfine field distribution from the same figure thus giving an idea of the amplitude of the $\mu$SR asymmetry oscillations.
} 
\label{fig-main} 
\end{figure}

The commensurate diagonal grid modulation was important to test,
because it implies a larger number of inequivalent nuclear sites, in comparison to the commensurate grid considered by Hunt {\it et al}.
However, as shown in Fig.~\ref{fig-main}(a), theoretical NQR spectrum corresponding to the diagonal commensurate grid modulation still has  too much structure in disagreement with experiment.


In the case of {\it incommensurate} periodic grid modulation not considered by Hunt {\it et al}., the distribution of local fields was obtained in the following way:

Grid superstructure is incommensurate, when the period of modulation along each of the two lattice directions is an irrational number [in the units of lattice period]. In this case, the two modulation phases are ergodic on the lattice, which means that, when taken modulo $2\pi$, both of them uniformly fill interval $(0, 2\pi)$. This implies that the distribution of the hyperfine fields can be computed from the modulation function (\ref{Hxy1}) by assuming that spins uniformly fill the space instead of being attached to a discrete lattice.

The resulting distribution function can be found analytically. For $H\leq H_m$, it has form
\begin{equation}
P(H) = \frac{F \left(
\hbox{Arccos}\left(\frac{H}{H_m}\right), \frac{1}{1-\left(\frac{H}{H_m}\right)^2}
\right)}{H_m \sqrt{1-\left(\frac{H}{H_m}\right)^2}}
,
\label{PH}
\end{equation}
where function $F(\xi, \eta)$ is the elliptic integral of the first kind:
\begin{equation}
F(\xi, \eta) = \int_0^{\xi} \frac{du}{\sqrt{1 - \eta \hbox{sin}^2 u}}.
\label{F}
\end{equation}
Here $\xi$ and $\eta$ are some function arguments, and $u$ is an integration variable.
For $H > H_m$, $P(H) = 0$.

In order to generate a specific distribution $P(H)$, $\delta=\frac{1}{8}$ was substituted into Eq.(\ref{Hm}), which gives $H_m = 3.16$~T. Strictly speaking $\delta = \frac{1}{8}$ corresponds to a commensurate modulation but the value of the small irrational number, which should be subtracted from $\frac{1}{8}$  makes very little difference, once the incommensurate modulation sets in and formula (\ref{PH}) becomes applicable.

The shape of $P(H)$ is shown in Fig.~\ref{fig-main}(b). It has a Van Hove singularity at $H=0$, which is associated with the fact that $H(x,y)$ given by Eq.(\ref{Hxy1}) has a saddle point at $x=0$, $y=0$. This singularity, however, does not lead to a pronounced feature in the NQR spectrum, because, at small values of $H$, the transition frequencies are dominated by the value of $\nu_Q$, and, therefore, the NQR spectrum is shaped by the non-singular distribution of $\nu_Q$. In contrast, the step feature of $P(H)$ at $H=H_m$ falls in the range, where the value of H begins controlling the values of transition frequencies. This leads to the [rounded] steps in the NQR spectrum at frequencies corresponding to $H=H_m$ (see Fig.~\ref{fig-freq}).

Overall, as shown in Fig.~\ref{fig-main}(b), the incommensurate grid modulation gives a better agreement with experiment in comparison with the commensurate one, but still with features in the form of steps absent in experiment. (A closer examination of experimental spectrum, in fact, reveals the hints of steps in the positions expected theoretically.)

In cuprates, there is no compelling reason to expect a perfectly uniform incommensurate spin density wave (1D or 2D). Instead, one should rather expect incommensurate periodicity only on average with local pattern distorted either by disorder or by the McMillan type of discommensurations in the commensurate superstructure\cite{McMillan-76}.  A straightforward way to incorporate both effects is to broaden the hyperfine field distribution obtained for the perfect incommensurate modulation.

As shown in Fig.~\ref{fig-main}(c), only a minor degree of broadening --- 10~per~cent of the original distribution width --- is sufficient to eliminate the essential disagreement between the theory and the experiment. The remaining disagreement should be well within the uncertainty of the theoretical scheme and the experiment (see the discussed in the next section).

The distribution of hyperfine fields in Fig.~\ref{fig-main}(c) was obtained by simple convolution  of the distribution shown in Fig.~\ref{fig-main}(b) with a Gaussian exp$-\frac{H^2}{2 \sigma_d^2}$, where $\sigma_d = 0.1 H_m$. This uniform broadening was chosen over the one proportional to $H$ [as done in Ref.\cite{Hunt-etal-01} and for Fig.~\ref{fig-main}(a)] just for the sake of simplicity. In general, one should expect $\sigma_d$ to have both $H$-independent and $H$-dependent terms. However, for the present purpose, the number of significance is $\sigma_d(H_m)$, while other differences in the choice of $\sigma_d(H)$ lead only to minute differences in the NQR spectra.

Ten per cent broadening is a very reasonable value theoretically. It is expected, for example, in the case, where a grid line fluctuates between two adjacent diagonals on the lattice, which leads to the fluctuations in the size of the grid supercell by $a/\sqrt{2}$. Given the approximate size of the supercell $4 a \sqrt{2}$, the above fluctuation amounts to 12.5 per cent fluctuation in the size of the supercell, which, in turn, can easily lead to a 10 per cent fluctuation in the value of $H_m$ within the supercell.

Importantly, the amplitude of spin modulation
used for the present calculation ($ \mu_0 = 0.3 \mu_B $ ) is consistent with the frequency of the $\mu$SR asymmetry oscillations\cite{Nachumi-etal-98}. One can further observe that the shape of the distribution of hyperfine fields seen by nuclear spins should not be too different from the shape of the distribution of magnetic fields experienced by muons. (Simulations in Ref.\cite{Nachumi-etal-98} indeed suggest so away from the close vicinity to the zero value of the field.) Therefore, by Fourier transforming the distribution of the hyperfine fields, one can obtain a rough estimate of the amplitude of the $\mu$SR asymmetry oscillations. Such a Fourier transform is shown in the inset of the right panel in Fig.~\ref{fig-main}(c). It exhibits clear
oscillations with the amplitude roughly consistent with but smaller than the $\mu$SR observations. (The frequency of $\mu$SR asymmetry oscillations should, of course be rescaled with
the strength of muon-electron magnetic coupling.)  It should be mentioned though, that, in the absense of NQR data, the the smallness of the oscillation amplitude was considered by Kojima {\it et al.} \cite{Kojima-etal-00} as an argument against the grid interpretation.

To summarize, none of the propositions considered by Hunt {\it et al}.\cite{Hunt-etal-01} came close to the kind of simultaneous agreement with the NQR and $\mu$SR results obtained in this section for the disordered incommensurate grid modulation.

\section{Discussion}
\label{discussion}

The significance of the findings presented in the previous section should be viewed in the context of uncertainties of the theoretical framework, within which these findings were obtained.

An important quantitative check of that framework is the fact, that it correctly predicts the experimental position of the 31~MHz peak. It is the check of the correctness of Hamiltonian (\ref{H}). More specifically, since the value of $\nu_Q$ is known from the high-temperature measurements, the above agreement amounts to a combined consistency check of the assumptions that the Hamiltonian is axially symmetric and that local hyperfine fields are parallel to the CuO$_2$ planes.

One uncertainty of the treatment in Section~\ref{computational} is associated with the assumption of completely random distribution of A- and B-type nuclei in the inhomogeneous background.
One can object,  that charged grid lines are attracted to dopant atoms, and hence the nuclei of B-type (attributed to the vicinity of dopant atoms) see quite different environment than A-type nuclei\cite{Ofer-etal-06}. It may happen, for example, that small amplitude fluctuations of grid lines wipe out the signal from B-nuclei, while leaving a significant signal coming from A-nuclei.  It turns out, however, that, if only A-type (majority) nuclei are included in the calculation for the disorder broadened incommensurate grid then the agreement between the theory and the experiment becomes even better than the one shown in Fig.~\ref{fig-main}(c).

A related difficulty is that the charge modulation, as it becomes stabilized, creates extra electric field gradient, which can lead to an extra broadening of the distribution of $\nu_Q$. A priori, the relative value of this broadening in comparison with the one due to the intrinsic factors (such as charged dopants or crystal imperfections) is difficult to estimate.  However, it is also difficult to see, why the modulation-induced broadening would be much larger than the ``intrinsic'' one, known from the width of the high temperature NQR spectra.  If it is not much larger, then the related uncertainty is not very significant.

Another important element missing from the description in Section~\ref{computational} is what happens in the system as a result of the $\pi/2$ and $\pi$ radio-frequency (rf) pulses of the spin echo sequence. The computational procedure of Section~\ref{computational} assumes that the $\pi/2$ and $\pi$ pulse conditions are perfectly satisfied for  every transition of every spin. In reality, this cannot be done simultaneously for all spins and all transitons, because (i) in powder samples (used in Ref.\cite{Hunt-etal-01}), the polarization of the rf field has a random orientation with respect to the lattice directions; and (ii) even, if it were a single crystal, the rf field at a given frequency affects simultaneously several NQR transitions and couples to those transitions through different values of matrix elements (which typically vary by a factor of 2). The difference in the values of matrix elements implies that the pulse conditions cannot be fulfilled simultaneously for every transition.

Even, if one were able to manipulate each transition individually, the use of weighing factor (\ref{Pnn}) would be doubtful. Such a factor is standard and appropriate, when the resonance is detected by the steady state absorption technique. In the pulsed experiments, however,  one power of the matrix element 
$| \langle n^{\prime} | \mathbf I  \cdot \mathbf H_1 |n \rangle|$ 
is absorbed into the duration of the rf pulses, which means that, if the $\pi/2$ and $\pi$ conditions are perfectly satisfied, one should expect a weighing factor equal to the above matrix element and not to its square as Eq.(\ref{Pnn}) suggests. When the pulse conditions are not perfectly satisfied, higher powers of that matrix element may contribute to the weighing factor, but even then, the use of the factor (\ref{Pnn}) would likely amount to a drastic oversimplification.
Expression~(\ref{Pnn}) was used in Section~\ref{computational} only to make the results directly comparable to the calculations of Hunt {\it et al}. 

Yet another uncertainty of this study is the lack of theory describing the wipeout of the NQR signals on the timescale of rf pulses.  In experiment of Ref.~\cite{Hunt-etal-01}, the duration of the rf pulses was chosen to be the one that maximizes the spin echo response. The wipeout effect would suppress the intensity of the echo, if the $\pi/2$ and $\pi$ conditions required too long pulses, and, therefore, could factor into the experimental definition of the pulse conditions. 

All unknown factors mentioned above can add up to a quantitatively significant weighing factor on the top of the theoretical calculation described in Section.~\ref{computational}. This extra factor, however, is not expected to have strong frequency dependence and, therefore, would not mask the presence of the multiple peaks associated either with discrete or with singular distributions arising from 1D modulations of hyperfine fields.

It is, therefore, the absence of the extra peaks beyond the one at 31~MHz that is the best indication against 1D stripe scenarios. 

The non-observation of multiple NQR peaks also constitutes an indication against spin spirals\cite{Sushkov-etal-05} in 1/8-doped lanthanum cuprates. If the Cu spins are indeed parallel to the CuO$_2$ plane, and if the NQR tensor is axially symmetric with respect to the $c$-axis,
then the spiral arrangement of spins implies that all nuclear sites are equivalent, which should lead to  sharp multiple NQR lines. If any of the above two assumptions is incorrect then NQR spectrum should still exhibit Van Hove singularities similar to those shown in Fig.\ref{fig-hunt}(d).

\section{Conclusions}
\label{conclusions}

The main message of this paper is that the absence of multiple peaks in the NQR spectra of Hunt et al. can be easily explained on the basis of slightly disordered incommensurate grid modulation. The degree of disorder required for this interpretation entails approximately 10 per cent broadening of the non-disordered NQR line, which would be consistent with the fact that the signatures of spin modulations are observable by $\mu$SR. At the same time, all alternative interpretations of the above spectra in terms of 1D spin modulations encounter serious difficulties. It is, therefore, concluded that the combined $\mu$SR/NQR data favor the the 2D grid modulation over the 1D stripe or spiral modulation static or dynamic.

{\it Note added:} After the present manuscript was submitted, this author became aware of the preprint by Christensen {\it et al.}\cite{Christensen-etal-07}, reporting significant new experimental results of magnetic neutron scattering in La$_{1.48}$Nd$_{0.4}$Sr$_{0.12}$CuO$_4$. These results were interpreted by the above authors as strongly supporting to the 1D stipe interpretation.  The present author has expressed his reservations on this matter in Ref.\cite{Fine-vortices-prb07}.

\section*{Acknowledgments}

The author is grateful to C.P.Slichter and J. Haase for drawing his attention to the NQR experiment of Hunt {\it et al}., to T. Imai for the detailed discussion of that experiment, and to T. Egami for discussions of this work.

\appendix*

\section{}

Figure~\ref{fig-peaks} shows relative peak intensities for charge modulation (\ref{rho}). These  intensities were obtained by the direct expansion of modulation (\ref{rho}) [with $a=1$] into Fourier harmonics:

\begin{widetext}
\begin{eqnarray}
\rho(x,y) &= {1 \over 256} \text{{\huge \{}}  &
70
-56 \  \text{cos}\left[\frac{\pi  x}{2}\right]
-56 \  \text{cos}\left[\frac{\pi  y}{2}\right]
+28 \  \text{cos}\left[\frac{\pi x}{2}+\frac{\pi  y}{2}\right]
+28 \  \text{cos}\left[\frac{\pi x}{2}-\frac{\pi  y}{2}\right]
\nonumber
\\
&&
+21 \  \text{cos}\left[\frac{\pi  x}{4}+\frac{\pi  y}{4}\right]
+21 \  \text{cos}\left[\frac{\pi  x}{4}-\frac{\pi  y}{4}\right]
\nonumber
\\
&&
-14 \  \text{cos}\left[\frac{3 \pi  x}{4}+\frac{\pi  y}{4}\right]
-14 \  \text{cos}\left[\frac{\pi  x}{4}+\frac{3 \pi  y}{4}\right]
-14 \  \text{cos}\left[\frac{\pi  x}{4}-\frac{3 \pi  y}{4}\right]
-14 \  \text{cos}\left[\frac{3 \pi  x}{4}-\frac{\pi  y}{4}\right]
\nonumber
\\
&&
+ \left( 
8 \  \text{cos}\left[\frac{3 \pi  x}{4}+\frac{3 \pi  y}{4}\right]
-\text{0.5$\grave{ }$}  \ \text{cos}\left[\frac{3 \pi  x}{4}-\frac{5\pi  y}{4}\right]
-\text{0.5$\grave{ }$}  \ \text{cos}\left[\frac{5 \pi  x}{4}-\frac{3 \pi  y}{4}\right]
\right)
\nonumber
\\
&&
+ \left( 
8 \  \text{cos}\left[\frac{3\pi  x}{4}-\frac{3 \pi  y}{4}\right]
-\text{0.5$\grave{}$} \  \text{cos}\left[\frac{5 \pi  x}{4}+\frac{3 \pi  y}{4}\right]
-\text{0.5$\grave{ }$} \  \text{cos}\left[\frac{3 \pi  x}{4}+\frac{5 \pi  y}{4}\right]
\right)
\nonumber
\\
&&
-4 \  \text{cos}\left[\frac{\pi  x}{2}+\pi  y\right]
-4 \  \text{cos}\left[\pi  x + \frac{\pi  y}{2}\right]
-4 \  \text{cos}\left[\frac{\pi  x}{2}-\pi  y\right]
-4 \  \text{cos}\left[\pi  x - \frac{\pi  y}{2}\right]
\nonumber
\\
&&
+ \text{cos}\left[\pi  x + \pi  y\right]
+ \text{cos}\left[\pi  x - \pi  y\right]
\hbox{{\huge \}}}
\label{expansion}
\end{eqnarray}

The terms grouped in parentheses (...) give the same peak positions in the first Brillouin zone. The relative peak intensities exhibited in Fig.~\ref{fig-peaks} are the squares of the Fourier amplitudes appearing in the above expansion. 

The following examples illustrate that the leading character of the peak at 
$(\pm \frac{\pi}{2}, 0) $ and $(0, \pm \frac{\pi}{2 })$ depends only  on the presence of the sin$^2$ factors, which suppress the charge density near the grid nodes.

\begin{equation}
\rho(x,y) = 
\left[ \hbox{cos} \frac{\pi(x+y)}{8} \right]^2 + 
\left[ \hbox{cos} \frac{\pi(-x+y)}{8} \right]^2 \  
=
{1 \over 2} \left\{2 + \text{cos}\left[\frac{\pi  x}{4}+\frac{\pi  y}{4}\right] + 
                      \text{cos}\left[\frac{\pi  x}{4}-\frac{\pi  y}{4}\right] \right\} 
\label{exp1}
\end{equation}

\begin{eqnarray}
\nonumber
\rho(x,y) &=& \left[ \hbox{cos} \frac{\pi(x+y)}{8} \right]^2 \  
\left[ \hbox{sin} \frac{\pi(-x+y)}{8} \right]^2
 + \ \left[ \hbox{cos} \frac{\pi(-x+y)}{8} \right]^2 \ 
\left[ \hbox{sin} \frac{\pi(x+y)}{8}\right]^2 
\\ 
&=&
{1 \over 4} \left\{  2 - \text{cos}\left[\frac{\pi  x}{2}\right] 
- \text{cos}\left[\frac{\pi  y}{2}\right]  \right\}
\label{exp2}
\end{eqnarray}

\begin{eqnarray}
\nonumber
\rho(x,y) &=& \left[ \hbox{cos} \frac{\pi(x+y)}{8} \right]^4 \  
\left[ \hbox{sin} \frac{\pi(-x+y)}{8} \right]^2
 + \ \left[ \hbox{cos} \frac{\pi(-x+y)}{8} \right]^4 \ 
\left[ \hbox{sin} \frac{\pi(x+y)}{8}\right]^2 
\\ 
&= {1 \over 32} \text{{\huge \{}}  &
12 
- 8 \  \text{cos}\left[\frac{\pi  x}{2}\right] 
- 8 \  \text{cos}\left[\frac{\pi  y}{2}\right]  
+ 2 \  \text{cos}\left[\frac{\pi x}{2}+\frac{\pi  y}{2}\right]
+ 2 \  \text{cos}\left[\frac{\pi x}{2}-\frac{\pi  y}{2}\right]
\nonumber
\\
&&
+ 2 \  \text{cos}\left[\frac{\pi  x}{4}+\frac{\pi  y}{4}\right]
+ 2 \  \text{cos}\left[\frac{\pi  x}{4}-\frac{\pi  y}{4}\right]
\nonumber
\\
&&
-  \  \text{cos}\left[\frac{3 \pi  x}{4}+\frac{\pi  y}{4}\right]
-  \  \text{cos}\left[\frac{\pi  x}{4}+\frac{3 \pi  y}{4}\right]
-  \  \text{cos}\left[\frac{\pi  x}{4}-\frac{3 \pi  y}{4}\right]
-  \  \text{cos}\left[\frac{3 \pi  x}{4}-\frac{\pi  y}{4}\right]
\hbox{{\huge \}}}
\label{exp3}
\end{eqnarray}

\end{widetext}


\begin{thebibliography}{49}
\expandafter\ifx\csname natexlab\endcsname\relax\def\natexlab#1{#1}\fi
\expandafter\ifx\csname bibnamefont\endcsname\relax
  \def\bibnamefont#1{#1}\fi
\expandafter\ifx\csname bibfnamefont\endcsname\relax
  \def\bibfnamefont#1{#1}\fi
\expandafter\ifx\csname citenamefont\endcsname\relax
  \def\citenamefont#1{#1}\fi
\expandafter\ifx\csname url\endcsname\relax
  \def\url#1{\texttt{#1}}\fi
\expandafter\ifx\csname urlprefix\endcsname\relax\def\urlprefix{URL }\fi
\providecommand{\bibinfo}[2]{#2}
\providecommand{\eprint}[2][]{\url{#2}}

\bibitem[{\citenamefont{Tranquada et~al.}(1995)\citenamefont{Tranquada,
  Sternlieb, Axe, Nakamura, and Uchida}}]{Tranquada-etal-95}
\bibinfo{author}{\bibfnamefont{J.~M.} \bibnamefont{Tranquada}},
  \bibinfo{author}{\bibfnamefont{B.~J.} \bibnamefont{Sternlieb}},
  \bibinfo{author}{\bibfnamefont{J.~D.} \bibnamefont{Axe}},
  \bibinfo{author}{\bibfnamefont{Y.}~\bibnamefont{Nakamura}}, \bibnamefont{and}
  \bibinfo{author}{\bibfnamefont{S.}~\bibnamefont{Uchida}},
  \bibinfo{journal}{Nature} \textbf{\bibinfo{volume}{375}},
  \bibinfo{pages}{561} (\bibinfo{year}{1995}).

\bibitem[{\citenamefont{Zaanen and Gunnarsson}(1989)}]{ZG}
\bibinfo{author}{\bibfnamefont{J.}~\bibnamefont{Zaanen}} \bibnamefont{and}
  \bibinfo{author}{\bibfnamefont{O.}~\bibnamefont{Gunnarsson}},
  \bibinfo{journal}{Phys. Rev. B} \textbf{\bibinfo{volume}{40}},
  \bibinfo{pages}{7391} (\bibinfo{year}{1989}).

\bibitem[{\citenamefont{Kato et~al.}(1990)\citenamefont{Kato, Machida,
  Nakanishi, and Fujita}}]{Kato-etal-90}
\bibinfo{author}{\bibfnamefont{M.}~\bibnamefont{Kato}},
  \bibinfo{author}{\bibfnamefont{K.}~\bibnamefont{Machida}},
  \bibinfo{author}{\bibfnamefont{H.}~\bibnamefont{Nakanishi}},
  \bibnamefont{and} \bibinfo{author}{\bibfnamefont{M.}~\bibnamefont{Fujita}},
  \bibinfo{journal}{J. Phys. Soc. Jpn} \textbf{\bibinfo{volume}{59}},
  \bibinfo{pages}{1047} (\bibinfo{year}{1990}).

\bibitem[{\citenamefont{Tranquada et~al.}(1999)\citenamefont{Tranquada,
  Ichikawa, Kakurai, and Uchida}}]{Tranquada-etal-99}
\bibinfo{author}{\bibfnamefont{J.~M.} \bibnamefont{Tranquada}},
  \bibinfo{author}{\bibfnamefont{N.}~\bibnamefont{Ichikawa}},
  \bibinfo{author}{\bibfnamefont{K.}~\bibnamefont{Kakurai}}, \bibnamefont{and}
  \bibinfo{author}{\bibfnamefont{S.}~\bibnamefont{Uchida}},
  \bibinfo{journal}{J. Phys. Chem. Solids} \textbf{\bibinfo{volume}{60}},
  \bibinfo{pages}{1019} (\bibinfo{year}{1999}).

\bibitem[{\citenamefont{Seibold and Grilli}(2001)}]{SG}
\bibinfo{author}{\bibfnamefont{G.}~\bibnamefont{Seibold}} \bibnamefont{and}
  \bibinfo{author}{\bibfnamefont{M.}~\bibnamefont{Grilli}},
  \bibinfo{journal}{Phys. Rev. B} \textbf{\bibinfo{volume}{63}},
  \bibinfo{pages}{224505} (\bibinfo{year}{2001}).

\bibitem[{\citenamefont{Fine}(2004{\natexlab{a}})}]{Fine-hitc-prb04}
\bibinfo{author}{\bibfnamefont{B.~V.} \bibnamefont{Fine}},
  \bibinfo{journal}{Phys. Rev. B} \textbf{\bibinfo{volume}{70}},
  \bibinfo{pages}{224508} (\bibinfo{year}{2004}{\natexlab{a}}),
  \bibinfo{note}{[eprint cond-mat/0308428, August 2003]}.

\bibitem[{\citenamefont{Fine}(2004{\natexlab{b}})}]{Fine-Miami-04}
\bibinfo{author}{\bibfnamefont{B.~V.} \bibnamefont{Fine}}, in
  \emph{\bibinfo{booktitle}{New Challenges in Superconductivity: Experimental
  Advances and Emerging Theories, Proceedings of NATO Advanced Research
  Workshop, 11-14 January 2004}}, edited by
  \bibinfo{editor}{\bibfnamefont{J.}~\bibnamefont{Ashkenazi}},
  \bibinfo{editor}{\bibfnamefont{M.}~\bibnamefont{Eremin}},
  \bibinfo{editor}{\bibfnamefont{J.~L.} \bibnamefont{Cohn}},
  \bibinfo{editor}{\bibfnamefont{I.}~\bibnamefont{Eremin}},
  \bibinfo{editor}{\bibfnamefont{D.}~\bibnamefont{Manske}},
  \bibinfo{editor}{\bibfnamefont{D.}~\bibnamefont{Pavuna}}, \bibnamefont{and}
  \bibinfo{editor}{\bibfnamefont{F.}~\bibnamefont{Zuo}}
  (\bibinfo{publisher}{Kluwer Academic Publishers},
  \bibinfo{year}{2004}{\natexlab{b}}), pp. \bibinfo{pages}{159--164},
  \bibinfo{note}{[e-print cond-mat/0404488]}.

\bibitem[{\citenamefont{Vershinin et~al.}(2004)\citenamefont{Vershinin, Misra,
  Ono, Abe, Ando, and Yazdani}}]{Vershinin-etal-04}
\bibinfo{author}{\bibfnamefont{M.}~\bibnamefont{Vershinin}},
  \bibinfo{author}{\bibfnamefont{S.}~\bibnamefont{Misra}},
  \bibinfo{author}{\bibfnamefont{S.}~\bibnamefont{Ono}},
  \bibinfo{author}{\bibfnamefont{Y.}~\bibnamefont{Abe}},
  \bibinfo{author}{\bibfnamefont{Y.}~\bibnamefont{Ando}}, \bibnamefont{and}
  \bibinfo{author}{\bibfnamefont{A.}~\bibnamefont{Yazdani}},
  \bibinfo{journal}{Science} \textbf{\bibinfo{volume}{303}},
  \bibinfo{pages}{1995} (\bibinfo{year}{2004}).

\bibitem[{\citenamefont{Hanaguri et~al.}(2004)\citenamefont{Hanaguri, Lupien,
  Kohsaka, Lee, Azuma, Takano, Takagi, and Davis}}]{Hanaguri-etal-04}
\bibinfo{author}{\bibfnamefont{T.}~\bibnamefont{Hanaguri}},
  \bibinfo{author}{\bibfnamefont{C.}~\bibnamefont{Lupien}},
  \bibinfo{author}{\bibfnamefont{Y.}~\bibnamefont{Kohsaka}},
  \bibinfo{author}{\bibfnamefont{D.~H.} \bibnamefont{Lee}},
  \bibinfo{author}{\bibfnamefont{M.}~\bibnamefont{Azuma}},
  \bibinfo{author}{\bibfnamefont{M.}~\bibnamefont{Takano}},
  \bibinfo{author}{\bibfnamefont{H.}~\bibnamefont{Takagi}}, \bibnamefont{and}
  \bibinfo{author}{\bibfnamefont{J.~C.} \bibnamefont{Davis}},
  \bibinfo{journal}{Nature} \textbf{\bibinfo{volume}{430}},
  \bibinfo{pages}{1001} (\bibinfo{year}{2004}).

\bibitem[{\citenamefont{Hinkov et~al.}(2004)\citenamefont{Hinkov,
  S.Pailh\`{e}s, Bourges, Sidis, Ivanov, Kulakov, Lin, Chen, Bernhard, and
  Keimer}}]{Hinkov-etal-04}
\bibinfo{author}{\bibfnamefont{V.}~\bibnamefont{Hinkov}},
  \bibinfo{author}{\bibnamefont{S.Pailh\`{e}s}},
  \bibinfo{author}{\bibfnamefont{P.}~\bibnamefont{Bourges}},
  \bibinfo{author}{\bibfnamefont{Y.}~\bibnamefont{Sidis}},
  \bibinfo{author}{\bibfnamefont{A.}~\bibnamefont{Ivanov}},
  \bibinfo{author}{\bibfnamefont{A.}~\bibnamefont{Kulakov}},
  \bibinfo{author}{\bibfnamefont{C.~T.} \bibnamefont{Lin}},
  \bibinfo{author}{\bibfnamefont{D.~P.} \bibnamefont{Chen}},
  \bibinfo{author}{\bibfnamefont{C.}~\bibnamefont{Bernhard}}, \bibnamefont{and}
  \bibinfo{author}{\bibfnamefont{B.}~\bibnamefont{Keimer}},
  \bibinfo{journal}{Nature} \textbf{\bibinfo{volume}{430}},
  \bibinfo{pages}{650} (\bibinfo{year}{2004}).

\bibitem[{\citenamefont{Yao et~al.}(2006)\citenamefont{Yao, Carlson, and
  Campbell}}]{Yao-etal-06}
\bibinfo{author}{\bibfnamefont{D.~X.} \bibnamefont{Yao}},
  \bibinfo{author}{\bibfnamefont{E.~W.} \bibnamefont{Carlson}},
  \bibnamefont{and} \bibinfo{author}{\bibfnamefont{D.~K.}
  \bibnamefont{Campbell}}, \bibinfo{journal}{Phys. Rev. B}
  \textbf{\bibinfo{volume}{73}}, \bibinfo{pages}{224525}
  (\bibinfo{year}{2006}).

\bibitem[{\citenamefont{Annett et~al.}(1996)\citenamefont{Annett, Goldenfeld,
  and Leggett}}]{AGL}
\bibinfo{author}{\bibfnamefont{J.~F.} \bibnamefont{Annett}},
  \bibinfo{author}{\bibfnamefont{N.}~\bibnamefont{Goldenfeld}},
  \bibnamefont{and} \bibinfo{author}{\bibfnamefont{A.~J.}
  \bibnamefont{Leggett}}, in \emph{\bibinfo{booktitle}{Physical Properties of
  High Temperature Superconductors, Vol.5}}, edited by
  \bibinfo{editor}{\bibfnamefont{D.~M.} \bibnamefont{Ginsberg}}
  (\bibinfo{publisher}{World Scientific, Singapore}, \bibinfo{year}{1996}), p.
  \bibinfo{pages}{375}, \bibinfo{note}{eprint: cond-mat/9601060}.

\bibitem[{\citenamefont{Damascelli et~al.}(2003)\citenamefont{Damascelli,
  Hussain, and Shen}}]{DHS}
\bibinfo{author}{\bibfnamefont{A.}~\bibnamefont{Damascelli}},
  \bibinfo{author}{\bibfnamefont{Z.}~\bibnamefont{Hussain}}, \bibnamefont{and}
  \bibinfo{author}{\bibfnamefont{Z.~X.} \bibnamefont{Shen}},
  \bibinfo{journal}{Rev. Mod. Phys.} \textbf{\bibinfo{volume}{75}},
  \bibinfo{pages}{473} (\bibinfo{year}{2003}).

\bibitem[{\citenamefont{Hoffman
  et~al.}(2002{\natexlab{a}})\citenamefont{Hoffman, Hudson, Lang, Madhavan,
  Eisaki, Uchida, and Davis}}]{Hoffman-etal-02}
\bibinfo{author}{\bibfnamefont{J.~E.} \bibnamefont{Hoffman}},
  \bibinfo{author}{\bibfnamefont{E.~W.} \bibnamefont{Hudson}},
  \bibinfo{author}{\bibfnamefont{K.~M.} \bibnamefont{Lang}},
  \bibinfo{author}{\bibfnamefont{V.}~\bibnamefont{Madhavan}},
  \bibinfo{author}{\bibfnamefont{H.}~\bibnamefont{Eisaki}},
  \bibinfo{author}{\bibfnamefont{S.}~\bibnamefont{Uchida}}, \bibnamefont{and}
  \bibinfo{author}{\bibfnamefont{J.~C.} \bibnamefont{Davis}},
  \bibinfo{journal}{Science} \textbf{\bibinfo{volume}{295}},
  \bibinfo{pages}{466} (\bibinfo{year}{2002}{\natexlab{a}}).

\bibitem[{\citenamefont{Hoffman
  et~al.}(2002{\natexlab{b}})\citenamefont{Hoffman, McElroy, Lee, Lang, Eisaki,
  Uchida, and Davis}}]{Hoffman-etal-02A}
\bibinfo{author}{\bibfnamefont{J.~E.} \bibnamefont{Hoffman}},
  \bibinfo{author}{\bibfnamefont{K.}~\bibnamefont{McElroy}},
  \bibinfo{author}{\bibfnamefont{D.-H.} \bibnamefont{Lee}},
  \bibinfo{author}{\bibfnamefont{K.~M.} \bibnamefont{Lang}},
  \bibinfo{author}{\bibfnamefont{H.}~\bibnamefont{Eisaki}},
  \bibinfo{author}{\bibfnamefont{S.}~\bibnamefont{Uchida}}, \bibnamefont{and}
  \bibinfo{author}{\bibfnamefont{J.~C.} \bibnamefont{Davis}},
  \bibinfo{journal}{Science} \textbf{\bibinfo{volume}{297}},
  \bibinfo{pages}{1148} (\bibinfo{year}{2002}{\natexlab{b}}).

\bibitem[{\citenamefont{Howald et~al.}(2003)\citenamefont{Howald, Eisaki,
  Kaneko, Greven, and Kapitulnik}}]{Howald-etal-03}
\bibinfo{author}{\bibfnamefont{C.}~\bibnamefont{Howald}},
  \bibinfo{author}{\bibfnamefont{H.}~\bibnamefont{Eisaki}},
  \bibinfo{author}{\bibfnamefont{N.}~\bibnamefont{Kaneko}},
  \bibinfo{author}{\bibfnamefont{M.}~\bibnamefont{Greven}}, \bibnamefont{and}
  \bibinfo{author}{\bibfnamefont{A.}~\bibnamefont{Kapitulnik}},
  \bibinfo{journal}{Phys. Rev. B} \textbf{\bibinfo{volume}{67}},
  \bibinfo{pages}{014533} (\bibinfo{year}{2003}).

\bibitem[{\citenamefont{Kivelson et~al.}(2003)\citenamefont{Kivelson, Bindloss,
  Fradkin, Oganesyan, Tranquada, Kapitulnik, and Howald}}]{Kivelson-etal-03}
\bibinfo{author}{\bibfnamefont{S.~A.} \bibnamefont{Kivelson}},
  \bibinfo{author}{\bibfnamefont{I.~P.} \bibnamefont{Bindloss}},
  \bibinfo{author}{\bibfnamefont{E.}~\bibnamefont{Fradkin}},
  \bibinfo{author}{\bibfnamefont{V.}~\bibnamefont{Oganesyan}},
  \bibinfo{author}{\bibfnamefont{J.~M.} \bibnamefont{Tranquada}},
  \bibinfo{author}{\bibfnamefont{A.}~\bibnamefont{Kapitulnik}},
  \bibnamefont{and} \bibinfo{author}{\bibfnamefont{C.}~\bibnamefont{Howald}},
  \bibinfo{journal}{Rev. Mod. Phys} \textbf{\bibinfo{volume}{75}},
  \bibinfo{pages}{1201} (\bibinfo{year}{2003}).

\bibitem[{\citenamefont{Robertson et~al.}()\citenamefont{Robertson, Kivelson,
  Fradkin, Fang, and Kapitulnik}}]{Robertson-etal-06}
\bibinfo{author}{\bibfnamefont{J.~A.} \bibnamefont{Robertson}},
  \bibinfo{author}{\bibfnamefont{S.~A.} \bibnamefont{Kivelson}},
  \bibinfo{author}{\bibfnamefont{E.}~\bibnamefont{Fradkin}},
  \bibinfo{author}{\bibfnamefont{A.~C.} \bibnamefont{Fang}}, \bibnamefont{and}
  \bibinfo{author}{\bibfnamefont{A.}~\bibnamefont{Kapitulnik}}, \eprint{eprint:
  cond-mat/0602675}.

\bibitem[{\citenamefont{Stock et~al.}(2006)\citenamefont{Stock, Buyers, Yamani,
  Broholm, Chung, Tun, Liang, Bonn, Hardy, and Birgeneau}}]{Stock-etal-06}
\bibinfo{author}{\bibfnamefont{C.}~\bibnamefont{Stock}},
  \bibinfo{author}{\bibfnamefont{W.~J.~L.} \bibnamefont{Buyers}},
  \bibinfo{author}{\bibfnamefont{Z.}~\bibnamefont{Yamani}},
  \bibinfo{author}{\bibfnamefont{C.~L.} \bibnamefont{Broholm}},
  \bibinfo{author}{\bibfnamefont{J.}~\bibnamefont{Chung}},
  \bibinfo{author}{\bibfnamefont{Z.}~\bibnamefont{Tun}},
  \bibinfo{author}{\bibfnamefont{R.}~\bibnamefont{Liang}},
  \bibinfo{author}{\bibfnamefont{D.}~\bibnamefont{Bonn}},
  \bibinfo{author}{\bibfnamefont{W.~N.} \bibnamefont{Hardy}}, \bibnamefont{and}
  \bibinfo{author}{\bibfnamefont{R.~J.} \bibnamefont{Birgeneau}},
  \bibinfo{journal}{Phys. Rev. B} \textbf{\bibinfo{volume}{73}},
  \bibinfo{pages}{100504(R)} (\bibinfo{year}{2006}).

\bibitem[{\citenamefont{Tranquada et~al.}(1996)\citenamefont{Tranquada, Axe,
  Ichikawa, Nakamura, Uchida, and Nachumi}}]{Tranquada-etal-96}
\bibinfo{author}{\bibfnamefont{J.~M.} \bibnamefont{Tranquada}},
  \bibinfo{author}{\bibfnamefont{J.~D.} \bibnamefont{Axe}},
  \bibinfo{author}{\bibfnamefont{N.}~\bibnamefont{Ichikawa}},
  \bibinfo{author}{\bibfnamefont{Y.}~\bibnamefont{Nakamura}},
  \bibinfo{author}{\bibfnamefont{S.}~\bibnamefont{Uchida}}, \bibnamefont{and}
  \bibinfo{author}{\bibfnamefont{B.}~\bibnamefont{Nachumi}},
  \bibinfo{journal}{Phys. Rev. B} \textbf{\bibinfo{volume}{54}},
  \bibinfo{pages}{7489} (\bibinfo{year}{1996}).

\bibitem[{\citenamefont{Yamada et~al.}(1998)\citenamefont{Yamada, Lee,
  Kurahashi, Wada, Wakimoto, Ueki, Kimura, Endoh, Hosoya, Shirane
  et~al.}}]{Yamada-etal-98}
\bibinfo{author}{\bibfnamefont{K.}~\bibnamefont{Yamada}},
  \bibinfo{author}{\bibfnamefont{C.~H.} \bibnamefont{Lee}},
  \bibinfo{author}{\bibfnamefont{K.}~\bibnamefont{Kurahashi}},
  \bibinfo{author}{\bibfnamefont{J.}~\bibnamefont{Wada}},
  \bibinfo{author}{\bibfnamefont{S.}~\bibnamefont{Wakimoto}},
  \bibinfo{author}{\bibfnamefont{S.}~\bibnamefont{Ueki}},
  \bibinfo{author}{\bibfnamefont{H.}~\bibnamefont{Kimura}},
  \bibinfo{author}{\bibfnamefont{Y.}~\bibnamefont{Endoh}},
  \bibinfo{author}{\bibfnamefont{S.}~\bibnamefont{Hosoya}},
  \bibinfo{author}{\bibfnamefont{G.}~\bibnamefont{Shirane}},
  \bibnamefont{et~al.}, \bibinfo{journal}{Phys. Rev. B}
  \textbf{\bibinfo{volume}{57}}, \bibinfo{pages}{6165} (\bibinfo{year}{1998}).

\bibitem[{\citenamefont{Fujita et~al.}(2002)\citenamefont{Fujita, Goka, Yamada,
  and Matsuda}}]{Fujita-etal-02}
\bibinfo{author}{\bibfnamefont{M.}~\bibnamefont{Fujita}},
  \bibinfo{author}{\bibfnamefont{H.}~\bibnamefont{Goka}},
  \bibinfo{author}{\bibfnamefont{K.}~\bibnamefont{Yamada}}, \bibnamefont{and}
  \bibinfo{author}{\bibfnamefont{M.}~\bibnamefont{Matsuda}},
  \bibinfo{journal}{Phys. Rev. Lett.} \textbf{\bibinfo{volume}{88}},
  \bibinfo{pages}{167008} (\bibinfo{year}{2002}).

\bibitem[{\citenamefont{Fujita et~al.}(2004)\citenamefont{Fujita, Goka, Yamada,
  Tranquada, and Regnault}}]{Fujita-etal-04}
\bibinfo{author}{\bibfnamefont{M.}~\bibnamefont{Fujita}},
  \bibinfo{author}{\bibfnamefont{H.}~\bibnamefont{Goka}},
  \bibinfo{author}{\bibfnamefont{K.}~\bibnamefont{Yamada}},
  \bibinfo{author}{\bibfnamefont{J.~M.} \bibnamefont{Tranquada}},
  \bibnamefont{and} \bibinfo{author}{\bibfnamefont{L.~P.}
  \bibnamefont{Regnault}}, \bibinfo{journal}{Phys. Rev. B}
  \textbf{\bibinfo{volume}{70}}, \bibinfo{pages}{104517}
  (\bibinfo{year}{2004}).

\bibitem[{\citenamefont{v.~Zimmermann et~al.}(1998)\citenamefont{v.~Zimmermann,
  Vigliante, Niem{\"o}ller, Ichikawa, Frello, Madsen, Wochner, Uchida,
  Andersen, Tranquada et~al.}}]{Zimmermann-etal-98}
\bibinfo{author}{\bibfnamefont{M.}~\bibnamefont{v.~Zimmermann}},
  \bibinfo{author}{\bibfnamefont{A.}~\bibnamefont{Vigliante}},
  \bibinfo{author}{\bibfnamefont{T.}~\bibnamefont{Niem{\"o}ller}},
  \bibinfo{author}{\bibfnamefont{N.}~\bibnamefont{Ichikawa}},
  \bibinfo{author}{\bibfnamefont{T.}~\bibnamefont{Frello}},
  \bibinfo{author}{\bibfnamefont{J.}~\bibnamefont{Madsen}},
  \bibinfo{author}{\bibfnamefont{P.}~\bibnamefont{Wochner}},
  \bibinfo{author}{\bibfnamefont{S.}~\bibnamefont{Uchida}},
  \bibinfo{author}{\bibfnamefont{N.~H.} \bibnamefont{Andersen}},
  \bibinfo{author}{\bibfnamefont{J.~M.} \bibnamefont{Tranquada}},
  \bibnamefont{et~al.}, \bibinfo{journal}{Europhys. Lett.}
  \textbf{\bibinfo{volume}{41}}, \bibinfo{pages}{629} (\bibinfo{year}{1998}).

\bibitem[{\citenamefont{Hunt et~al.}(2001)\citenamefont{Hunt, Singer,
  Cederstr{\"o}m, and Imai}}]{Hunt-etal-01}
\bibinfo{author}{\bibfnamefont{A.~W.} \bibnamefont{Hunt}},
  \bibinfo{author}{\bibfnamefont{P.~M.} \bibnamefont{Singer}},
  \bibinfo{author}{\bibfnamefont{A.~F.} \bibnamefont{Cederstr{\"o}m}},
  \bibnamefont{and} \bibinfo{author}{\bibfnamefont{T.}~\bibnamefont{Imai}},
  \bibinfo{journal}{Phys. Rev. B} \textbf{\bibinfo{volume}{64}},
  \bibinfo{pages}{134525} (\bibinfo{year}{2001}).

\bibitem[{\citenamefont{Nachumi et~al.}(1998)\citenamefont{Nachumi, Fudamoto,
  Keren, Kojima, Larkin, Luke, Merrin, Tschernyshyov, Uemura, Ichikawa
  et~al.}}]{Nachumi-etal-98}
\bibinfo{author}{\bibfnamefont{B.}~\bibnamefont{Nachumi}},
  \bibinfo{author}{\bibfnamefont{Y.}~\bibnamefont{Fudamoto}},
  \bibinfo{author}{\bibfnamefont{A.}~\bibnamefont{Keren}},
  \bibinfo{author}{\bibfnamefont{K.~M.} \bibnamefont{Kojima}},
  \bibinfo{author}{\bibfnamefont{M.}~\bibnamefont{Larkin}},
  \bibinfo{author}{\bibfnamefont{G.~M.} \bibnamefont{Luke}},
  \bibinfo{author}{\bibfnamefont{J.}~\bibnamefont{Merrin}},
  \bibinfo{author}{\bibfnamefont{O.}~\bibnamefont{Tschernyshyov}},
  \bibinfo{author}{\bibfnamefont{Y.~J.} \bibnamefont{Uemura}},
  \bibinfo{author}{\bibfnamefont{N.}~\bibnamefont{Ichikawa}},
  \bibnamefont{et~al.}, \bibinfo{journal}{Phys. Rev. B}
  \textbf{\bibinfo{volume}{58}}, \bibinfo{pages}{8760} (\bibinfo{year}{1998}).

\bibitem[{\citenamefont{Shraiman and Siggia}(1988)}]{Shraiman-etal-88}
\bibinfo{author}{\bibfnamefont{B.~I.} \bibnamefont{Shraiman}} \bibnamefont{and}
  \bibinfo{author}{\bibfnamefont{E.~D.} \bibnamefont{Siggia}},
  \bibinfo{journal}{Phys. Rev. Lett.} \textbf{\bibinfo{volume}{61}},
  \bibinfo{pages}{467} (\bibinfo{year}{1988}).

\bibitem[{\citenamefont{Hasselmann et~al.}(2004)\citenamefont{Hasselmann,
  {Castro Neto}, and Smith}}]{Hasselmann-etal-04}
\bibinfo{author}{\bibfnamefont{N.}~\bibnamefont{Hasselmann}},
  \bibinfo{author}{\bibfnamefont{A.~H.} \bibnamefont{{Castro Neto}}},
  \bibnamefont{and} \bibinfo{author}{\bibfnamefont{C.~M.} \bibnamefont{Smith}},
  \bibinfo{journal}{Phys. Rev. B} \textbf{\bibinfo{volume}{69}},
  \bibinfo{pages}{014424} (\bibinfo{year}{2004}).

\bibitem[{\citenamefont{Sushkov and Kotov}(2005)}]{Sushkov-etal-05}
\bibinfo{author}{\bibfnamefont{O.~P.} \bibnamefont{Sushkov}} \bibnamefont{and}
  \bibinfo{author}{\bibfnamefont{V.~N.} \bibnamefont{Kotov}},
  \bibinfo{journal}{Phys. Rev. Lett.} \textbf{\bibinfo{volume}{94}},
  \bibinfo{pages}{097005} (\bibinfo{year}{2005}).

\bibitem[{\citenamefont{Abbamonte et~al.}(2006)\citenamefont{Abbamonte, Rusydi,
  Smadici, Gu, Sawatzky, and Feng}}]{Abbamonte-etal-06}
\bibinfo{author}{\bibfnamefont{P.}~\bibnamefont{Abbamonte}},
  \bibinfo{author}{\bibfnamefont{A.}~\bibnamefont{Rusydi}},
  \bibinfo{author}{\bibfnamefont{S.}~\bibnamefont{Smadici}},
  \bibinfo{author}{\bibfnamefont{G.~D.} \bibnamefont{Gu}},
  \bibinfo{author}{\bibfnamefont{G.~A.} \bibnamefont{Sawatzky}},
  \bibnamefont{and} \bibinfo{author}{\bibfnamefont{D.~L.} \bibnamefont{Feng}},
  \bibinfo{journal}{Nature Physics} \textbf{\bibinfo{volume}{1}},
  \bibinfo{pages}{155} (\bibinfo{year}{2006}).

\bibitem[{\citenamefont{Zachar et~al.}(1998)\citenamefont{Zachar, Kivelson, and
  Emery}}]{ZKE}
\bibinfo{author}{\bibfnamefont{O.}~\bibnamefont{Zachar}},
  \bibinfo{author}{\bibfnamefont{S.~A.} \bibnamefont{Kivelson}},
  \bibnamefont{and} \bibinfo{author}{\bibfnamefont{V.~J.} \bibnamefont{Emery}},
  \bibinfo{journal}{Phys. Rev. B.} \textbf{\bibinfo{volume}{57}},
  \bibinfo{pages}{1422} (\bibinfo{year}{1998}).

\bibitem[{\citenamefont{Hirota et~al.}(1998)\citenamefont{Hirota, Yamada,
  Tanaka, and Kojima}}]{Hirota-etal-98}
\bibinfo{author}{\bibfnamefont{K.}~\bibnamefont{Hirota}},
  \bibinfo{author}{\bibfnamefont{K.}~\bibnamefont{Yamada}},
  \bibinfo{author}{\bibfnamefont{I.}~\bibnamefont{Tanaka}}, \bibnamefont{and}
  \bibinfo{author}{\bibfnamefont{H.}~\bibnamefont{Kojima}},
  \bibinfo{journal}{Physica B} \textbf{\bibinfo{volume}{241-243}},
  \bibinfo{pages}{817} (\bibinfo{year}{1998}).

\bibitem[{\citenamefont{Kimura et~al.}(1999)\citenamefont{Kimura, Hirota,
  Matsushita, Yamada, Endoh, Lee, Majkrzak, Erwin, Shirane, Greven
  et~al.}}]{Kimura-etal-99}
\bibinfo{author}{\bibfnamefont{H.}~\bibnamefont{Kimura}},
  \bibinfo{author}{\bibfnamefont{K.}~\bibnamefont{Hirota}},
  \bibinfo{author}{\bibfnamefont{H.}~\bibnamefont{Matsushita}},
  \bibinfo{author}{\bibfnamefont{K.}~\bibnamefont{Yamada}},
  \bibinfo{author}{\bibfnamefont{Y.}~\bibnamefont{Endoh}},
  \bibinfo{author}{\bibfnamefont{S.}~\bibnamefont{Lee}},
  \bibinfo{author}{\bibfnamefont{C.~F.} \bibnamefont{Majkrzak}},
  \bibinfo{author}{\bibfnamefont{R.}~\bibnamefont{Erwin}},
  \bibinfo{author}{\bibfnamefont{G.}~\bibnamefont{Shirane}},
  \bibinfo{author}{\bibfnamefont{M.}~\bibnamefont{Greven}},
  \bibnamefont{et~al.}, \bibinfo{journal}{Phys. Rev. B}
  \textbf{\bibinfo{volume}{59}}, \bibinfo{pages}{6517} (\bibinfo{year}{1999}).

\bibitem[{\citenamefont{Hirota}(2001)}]{Hirota-01}
\bibinfo{author}{\bibfnamefont{K.}~\bibnamefont{Hirota}},
  \bibinfo{journal}{Physica C} \textbf{\bibinfo{volume}{357-360}},
  \bibinfo{pages}{61} (\bibinfo{year}{2001}).

\bibitem[{LTT()}]{LTTenergy}
\bibinfo{note}{This assumption is also consistent with the theoretical
  expectation that, the electronic energies involved in the formation of
  modulation patterns are much larger than the energy of the coupling between
  the resulting modulation and the lattice anisotropy.}

\bibitem[{\citenamefont{Hunt et~al.}(1999)\citenamefont{Hunt, Singer, Thurber,
  and Imai}}]{Hunt-etal-99}
\bibinfo{author}{\bibfnamefont{A.~W.} \bibnamefont{Hunt}},
  \bibinfo{author}{\bibfnamefont{P.~M.} \bibnamefont{Singer}},
  \bibinfo{author}{\bibfnamefont{K.~R.} \bibnamefont{Thurber}},
  \bibnamefont{and} \bibinfo{author}{\bibfnamefont{T.}~\bibnamefont{Imai}},
  \bibinfo{journal}{Phys. Rev. Lett.} \textbf{\bibinfo{volume}{82}},
  \bibinfo{pages}{4300} (\bibinfo{year}{1999}).

\bibitem[{\citenamefont{Curro et~al.}(2000)\citenamefont{Curro, Hammel, Suh,
  cker, B{\"u}chner, Ammerahl, and Revcolecschi}}]{Curro-etal-00}
\bibinfo{author}{\bibfnamefont{N.~J.} \bibnamefont{Curro}},
  \bibinfo{author}{\bibfnamefont{P.~C.} \bibnamefont{Hammel}},
  \bibinfo{author}{\bibfnamefont{B.~J.} \bibnamefont{Suh}},
  \bibinfo{author}{\bibfnamefont{M.~H.} \bibnamefont{cker}},
  \bibinfo{author}{\bibfnamefont{B.}~\bibnamefont{B{\"u}chner}},
  \bibinfo{author}{\bibfnamefont{U.}~\bibnamefont{Ammerahl}}, \bibnamefont{and}
  \bibinfo{author}{\bibfnamefont{A.}~\bibnamefont{Revcolecschi}},
  \bibinfo{journal}{Phys. Rev. Lett.} \textbf{\bibinfo{volume}{85}},
  \bibinfo{pages}{642} (\bibinfo{year}{2000}).

\bibitem[{\citenamefont{Suh et~al.}(2000)\citenamefont{Suh, Hammel, cker,
  B{\"u}chner, Ammerahl, and Revcolecschi}}]{Suh-etal-00}
\bibinfo{author}{\bibfnamefont{B.~J.} \bibnamefont{Suh}},
  \bibinfo{author}{\bibfnamefont{P.~C.} \bibnamefont{Hammel}},
  \bibinfo{author}{\bibfnamefont{M.~H.} \bibnamefont{cker}},
  \bibinfo{author}{\bibfnamefont{B.}~\bibnamefont{B{\"u}chner}},
  \bibinfo{author}{\bibfnamefont{U.}~\bibnamefont{Ammerahl}}, \bibnamefont{and}
  \bibinfo{author}{\bibfnamefont{A.}~\bibnamefont{Revcolecschi}},
  \bibinfo{journal}{Phys. Rev. B} \textbf{\bibinfo{volume}{61}},
  \bibinfo{pages}{R9265} (\bibinfo{year}{2000}).

\bibitem[{\citenamefont{Teitel'baum
  et~al.}(2000{\natexlab{a}})\citenamefont{Teitel'baum, B{\"u}chner, and
  de~Gronckel}}]{Teitelbaum-etal-00}
\bibinfo{author}{\bibfnamefont{G.~B.} \bibnamefont{Teitel'baum}},
  \bibinfo{author}{\bibfnamefont{B.}~\bibnamefont{B{\"u}chner}},
  \bibnamefont{and}
  \bibinfo{author}{\bibfnamefont{H.}~\bibnamefont{de~Gronckel}},
  \bibinfo{journal}{Phys. Rev. Lett.} \textbf{\bibinfo{volume}{84}},
  \bibinfo{pages}{2949} (\bibinfo{year}{2000}{\natexlab{a}}).

\bibitem[{\citenamefont{Teitel'baum
  et~al.}(2000{\natexlab{b}})\citenamefont{Teitel'baum, {Abu-Shiekah},
  Bakharev, and Brom}}]{Teitelbaum-etal-00A}
\bibinfo{author}{\bibfnamefont{G.~B.} \bibnamefont{Teitel'baum}},
  \bibinfo{author}{\bibfnamefont{I.~M.} \bibnamefont{{Abu-Shiekah}}},
  \bibinfo{author}{\bibfnamefont{O.}~\bibnamefont{Bakharev}}, \bibnamefont{and}
  \bibinfo{author}{\bibfnamefont{H.~B.} \bibnamefont{Brom}},
  \bibinfo{journal}{Phys. Rev. B} \textbf{\bibinfo{volume}{63}},
  \bibinfo{pages}{020507(R)} (\bibinfo{year}{2000}{\natexlab{b}}).

\bibitem[{\citenamefont{Haase et~al.}(2000)\citenamefont{Haase, Slichter,
  Stern, Milling, and Hinks}}]{Haase-etal-00}
\bibinfo{author}{\bibfnamefont{J.}~\bibnamefont{Haase}},
  \bibinfo{author}{\bibfnamefont{C.~P.} \bibnamefont{Slichter}},
  \bibinfo{author}{\bibfnamefont{R.}~\bibnamefont{Stern}},
  \bibinfo{author}{\bibfnamefont{C.~T.} \bibnamefont{Milling}},
  \bibnamefont{and} \bibinfo{author}{\bibfnamefont{D.~G.} \bibnamefont{Hinks}},
  \bibinfo{journal}{J. Supercond.} \textbf{\bibinfo{volume}{13}},
  \bibinfo{pages}{723} (\bibinfo{year}{2000}).

\bibitem[{\citenamefont{Haase et~al.}(2002)\citenamefont{Haase, Slichter, and
  Milling}}]{Haase-etal-02}
\bibinfo{author}{\bibfnamefont{J.}~\bibnamefont{Haase}},
  \bibinfo{author}{\bibfnamefont{C.~P.} \bibnamefont{Slichter}},
  \bibnamefont{and} \bibinfo{author}{\bibfnamefont{C.~T.}
  \bibnamefont{Milling}}, \bibinfo{journal}{J. Supercond.}
  \textbf{\bibinfo{volume}{15}}, \bibinfo{pages}{339} (\bibinfo{year}{2002}).

\bibitem[{\citenamefont{Imai et~al.}(1993)\citenamefont{Imai, Slichter,
  Yoshimura, and Kosuge}}]{Imai-etal-93}
\bibinfo{author}{\bibfnamefont{T.}~\bibnamefont{Imai}},
  \bibinfo{author}{\bibfnamefont{C.~P.} \bibnamefont{Slichter}},
  \bibinfo{author}{\bibfnamefont{K.}~\bibnamefont{Yoshimura}},
  \bibnamefont{and} \bibinfo{author}{\bibfnamefont{K.}~\bibnamefont{Kosuge}},
  \bibinfo{journal}{Phys. Rev. Lett.} \textbf{\bibinfo{volume}{70}},
  \bibinfo{pages}{1002} (\bibinfo{year}{1993}).

\bibitem[{\citenamefont{Vaknin et~al.}(1987)\citenamefont{Vaknin, Sinha,
  Moncton, Johnston, Newsam, Safinya, and King}}]{Vaknin-etal-87}
\bibinfo{author}{\bibfnamefont{D.}~\bibnamefont{Vaknin}},
  \bibinfo{author}{\bibfnamefont{S.~K.} \bibnamefont{Sinha}},
  \bibinfo{author}{\bibfnamefont{D.~E.} \bibnamefont{Moncton}},
  \bibinfo{author}{\bibfnamefont{D.~C.} \bibnamefont{Johnston}},
  \bibinfo{author}{\bibfnamefont{J.~M.} \bibnamefont{Newsam}},
  \bibinfo{author}{\bibfnamefont{C.~R.} \bibnamefont{Safinya}},
  \bibnamefont{and} \bibinfo{author}{\bibfnamefont{H.~E.} \bibnamefont{King}},
  \bibinfo{journal}{Phys. Rev. Lett.} \textbf{\bibinfo{volume}{58}},
  \bibinfo{pages}{2802} (\bibinfo{year}{1987}).

\bibitem[{\citenamefont{McMillan}(1976)}]{McMillan-76}
\bibinfo{author}{\bibfnamefont{W.~L.} \bibnamefont{McMillan}},
  \bibinfo{journal}{Phys. Rev. B} \textbf{\bibinfo{volume}{14}},
  \bibinfo{pages}{1496} (\bibinfo{year}{1976}).

\bibitem[{\citenamefont{Kojima et~al.}(2000)\citenamefont{Kojima, Eisaki,
  Uchida, Fudamoto, Gat, Kinkhabwala, Larkin, Luke, and
  Uemura}}]{Kojima-etal-00}
\bibinfo{author}{\bibfnamefont{K.~M.} \bibnamefont{Kojima}},
  \bibinfo{author}{\bibfnamefont{H.}~\bibnamefont{Eisaki}},
  \bibinfo{author}{\bibfnamefont{S.}~\bibnamefont{Uchida}},
  \bibinfo{author}{\bibfnamefont{Y.}~\bibnamefont{Fudamoto}},
  \bibinfo{author}{\bibfnamefont{I.~M.} \bibnamefont{Gat}},
  \bibinfo{author}{\bibfnamefont{A.}~\bibnamefont{Kinkhabwala}},
  \bibinfo{author}{\bibfnamefont{M.~I.} \bibnamefont{Larkin}},
  \bibinfo{author}{\bibfnamefont{G.~M.} \bibnamefont{Luke}}, \bibnamefont{and}
  \bibinfo{author}{\bibfnamefont{Y.~J.} \bibnamefont{Uemura}},
  \bibinfo{journal}{Physica B} \textbf{\bibinfo{volume}{289-290}},
  \bibinfo{pages}{343} (\bibinfo{year}{2000}).

\bibitem[{\citenamefont{Ofer et~al.}(2006)\citenamefont{Ofer, Levy, Kanigel,
  and Keren}}]{Ofer-etal-06}
\bibinfo{author}{\bibfnamefont{R.}~\bibnamefont{Ofer}},
  \bibinfo{author}{\bibfnamefont{S.}~\bibnamefont{Levy}},
  \bibinfo{author}{\bibfnamefont{A.}~\bibnamefont{Kanigel}}, \bibnamefont{and}
  \bibinfo{author}{\bibfnamefont{A.}~\bibnamefont{Keren}},
  \bibinfo{journal}{Phys. Rev. B} \textbf{\bibinfo{volume}{73}},
  \bibinfo{pages}{012503} (\bibinfo{year}{2006}).

\bibitem[{\citenamefont{Christensen et~al.}(2007)\citenamefont{Christensen,
  R{\o{}}nnow, Mesot, Ewings, Momono, Oda, Ido, Enderle, McMorrow, and
  Boothroyd}}]{Christensen-etal-07}
\bibinfo{author}{\bibfnamefont{N.~B.} \bibnamefont{Christensen}},
  \bibinfo{author}{\bibfnamefont{H.~M.} \bibnamefont{R{\o{}}nnow}},
  \bibinfo{author}{\bibfnamefont{J.}~\bibnamefont{Mesot}},
  \bibinfo{author}{\bibfnamefont{R.~A.} \bibnamefont{Ewings}},
  \bibinfo{author}{\bibfnamefont{N.}~\bibnamefont{Momono}},
  \bibinfo{author}{\bibfnamefont{M.}~\bibnamefont{Oda}},
  \bibinfo{author}{\bibfnamefont{M.}~\bibnamefont{Ido}},
  \bibinfo{author}{\bibfnamefont{M.}~\bibnamefont{Enderle}},
  \bibinfo{author}{\bibfnamefont{D.~F.} \bibnamefont{McMorrow}},
  \bibnamefont{and} \bibinfo{author}{\bibfnamefont{A.~T.}
  \bibnamefont{Boothroyd}}, \bibinfo{journal}{Phys. Rev. Lett.}
  \textbf{\bibinfo{volume}{98}}, \bibinfo{pages}{197003}
  (\bibinfo{year}{2007}), \eprint{eprint: cond-mat/0608204}.

\bibitem[{\citenamefont{Fine}(2007)}]{Fine-vortices-prb07}
\bibinfo{author}{\bibfnamefont{B.~V.} \bibnamefont{Fine}},
  \bibinfo{journal}{Phys. Rev. B} \textbf{\bibinfo{volume}{75}},
  \bibinfo{pages}{060504} (\bibinfo{year}{2007}), \eprint{eprint:
  cond-mat/0610748}.

\end{thebibliography}

\end{document}